\documentclass[structabstract]{aa}
\pdfoutput=1
\usepackage{graphicx}
\usepackage{txfonts}
\usepackage{natbib}
\usepackage{multirow}

\newcommand{\alphauj}{<\alpha_u>_j}
\newcommand{\alphapj}{<\alpha_p>_j}
\newcommand{\alphafj}{<\alpha_f>_j}
\newcommand{\alphanj}{<\alpha_n>_j}

\newcommand{\Fj}{<F(\lambda)>_j}

\newcommand{\Rj}{<R>_j}
\newcommand{\openj}{<\phi_{open}>_j}
\newcommand{\openjj}{<\phi_{open}>_{j+1}}

\newcommand{\facj}{<\phi_{f}>_j}
\newcommand{\netj}{<\phi_{n}>_j}
\newcommand{\actj}{<\phi_{act}>_j}

\newcommand{\ephj}{<\phi_{eph}>_j}

\newcommand{\uj}{<\phi_{u}>_j}
\newcommand{\pj}{<\phi_{p}>_j}

\bibpunct{(}{)}{;}{a}{}{,} 

\begin{document}

\title{Evolution of the solar irradiance during the Holocene}

\author{Luis Eduardo A. Vieira\inst{1,2}, Sami K. Solanki\inst{1,3}, Natalie A. Krivova\inst{1}, Ilya Usoskin\inst{4}}

\offprints{L.E.A. Vieira, \email{luis.vieira@cnrs-orleans.fr}}

\institute{Max-Planck-Institut f\"ur Sonnensystemforschung, Max-Planck-Str. 2, 37191 Katlenburg-Lindau, Germany \and Laboratoire de Physique et Chimie de l'Environnement et de l'Espace  (LPC2E/CNRS), 3A, Avenue de la Recherche, 45071 Orléans cedex 2, France  \and  School of Space Research, Kyung Hee University, Yongin, Gyeonggi, 446-701, Korea \and  Sodankyla Geophysical Observatory (Oulu Unit), POB 3000, Universiy of Oulu, Finland}

\date{Received  / Accepted }

\abstract { Long-term records of solar radiative output are vital for understanding solar variability and past climate change. Measurements of solar irradiance are available for only the last three decades, which calls for reconstructions of this quantity over longer time scales using suitable models.}{We present a physically consistent reconstruction of the total solar irradiance for the Holocene.} {We extend the SATIRE (Spectral And Total Irradiance REconstruction) models to estimate the evolution of the total (and partly spectral) solar irradiance over the Holocene. The basic assumption is that the variations of the solar irradiance are due to the evolution of the dark and bright magnetic features on the solar surface. The evolution of the decadally averaged magnetic flux is computed from decadal values of cosmogenic isotope concentrations recorded in natural archives employing a series of physics-based models connecting the processes from the modulation of the cosmic ray flux in the heliosphere to their record in natural archives. We then compute the total solar irradiance (TSI) as a linear combination of the $j^{th}$ and $j^{th}+1$ decadal values of the open magnetic flux. In order to evaluate the uncertainties due to the evolution of the Earth's magnetic dipole moment, we employ four  reconstructions of the open flux which are based on conceptually different paleomagnetic models.}{Reconstructions of the TSI over the Holocene, each valid for a different paleomagnetic time series, are presented. Our analysis suggests that major sources of uncertainty in the TSI in this model are the heritage of the uncertainty of the TSI since 1610 reconstructed from sunspot data and the uncertainty of the evolution of the Earth's magnetic dipole moment. The analysis of the distribution functions of the reconstructed irradiance for the last 3000 years, which is the period that the reconstructions overlap, indicates that the estimates based on the virtual axial dipole moment are significantly lower at earlier times than the reconstructions based on the virtual dipole moment. We also present a combined reconstruction, which represents our best estimate of total solar irradiance for any given time during the Holocene.}{We present the first physics-based reconstruction of the total solar irradiance over the Holocene, which will be of interest for studies of climate change over the last 11500 years. The reconstruction indicates that the decadally averaged total solar irradiance ranges over approximately 1.5 W/m$^2$ from grand maxima to grand minima.}

 \keywords{Sun: activity;  Sun: faculae, plages; Sun: surface magnetism;  Sun: solar-terrestrial relations; Sun: sunspots; Sun: UV radiation}

\authorrunning{Vieira, Solanki, Krivova, and Usoskin}

\titlerunning{Evolution of the TSI/SSI}

\maketitle


\section{Introduction}

Radiative energy input from the Sun is the main external source of heat to the Earth's coupled atmospheric/oceanic system \citep{hansen2000,haigh2001}.
Variations in this external source obviously affect the climate, although there is still considerable uncertainty and debate on the level
 of this effect \citep{haigh2001, hansen2002, hansen2005, ipcc2007,camp2007}. One way to disentangle the influence of the Sun on the climate is to look for its influence on climate in pre-industrial times, when the influence of man-made greenhouse gases can be neglected and the natural variations can be isolated.

Direct measurements of atmospheric and oceanic parameters during the last century suggest the occurrence of climate shifts on
 several time scales \citep{trenberth1990, zhang2006, ipcc2007}. Holocene climate proxies also suggest the occurrence of millennial scale climate variations \citep{rodbell1999, bond2001}. For example, the intensity and frequency of the El Nino and Southern Oscillation (ENSO), which is the most important oscillation mode of the atmosphere/ocean coupled system, seems to have undergone significant climatic shifts in the Holocene \citep{rodbell1999}, with strong ENSO events becoming established only about 5,000 years ago. Furthermore, Holocene climate proxies also suggest substantial correlations between atmospheric and oceanic conditions and solar variations, but the physical mechanisms are not understood yet \citep{bond2001, haigh2001,haigh2007, usoskin2005}.

With the current generation of general circulation models it is now possible to carry out long-term numerical simulations of the
 evolution of the Earth's climate system \citep{jungclaus2010, schmidt2010}.
Such studies require long time series of solar irradiance, ideally covering the whole Holocene.
Although reconstructions of irradiance since the Maunder Minimum are common
 \citep[e.g.][]{lean1995, solanki1999, fligge2000, wang2005, balmaceda2007, krivova2007,  krivova2010a, tapping2007, crouch2008}, on
 longer timescales few quantitative reconstructions are available due to the lack of sufficiently time-resolved solar data and
 appropriate techniques for reconstructing the irradiance when only data averaged over 10 years (i.e. roughly a solar cycle) are
 available \citep[but see][]{steinhilber2009}.
Here we compute the total (and to a limited extend the spectral) irradiance for the Holocene employing the reconstructions of the
 open flux and sunspot number obtained from the cosmogenic isotope \element[ ][14]{C}.
The motivation for this work came from a study carried out by \cite{balmaceda2007}, who compared the annual and cycle averages
 of the TSI, total and open magnetic fluxes, as computed using the model described by \cite{krivova2007}, and the observed sunspot number.
She found robust relationships between these quantities.
\cite{vieira2008agu} employed these relationships to make a preliminary computation of the total solar irradiance for the Holocene from
 the sunspot number and open magnetic flux estimated from \element[ ][14]{C}.
More recently, \cite{steinhilber2009} presented a reconstruction of the total solar irradiance for the Holocene based
 on the cosmogenic radionuclide \element[][10 ]{Be} measured in ice cores.
A shorter reconstruction has been presented by \cite{delaygue2010}.

In the present work we derive the relationship between solar cycle averaged open magnetic flux and total solar irradiance for the physics-based model of \cite{krivova2007, krivova2010a}. This is then used to reconstruct the irradiance throughout the Holocene based on \element[ ][14]{C} data. The reconstructed values lying in the telescope era are compared with the cycle averaged values obtained by \cite{krivova2010a}. The present model is so far unique in that it is based on a consistently derived, physics-based relationship between TSI and the magnetic flux, which incidentally does not correspond to a direct, instantaneous relationship as is often incorrectly assumed. {By following this approach the real physical processes are considered in all steps, but some parameters, which cannot be obtained directly from theory, are evaluated by fitting the model's output to observational data.}

We structure the paper as follows. In Sect. 2, we describe the physics-based model used to reconstruct the TSI and SSI, which is, in principle, identical to the models employed by \cite{krivova2007, krivova2010a} to calculate the solar irradiance on time scales from days to centuries, but adapted to work with heavily time-averaged data. In Sect. 3, the reconstruction of the TSI and SSI for the Holocene is presented and sources of uncertainty are discussed. Finally, we present the conclusions in Sect. 4.

\section{Model Description}

The primary source of information on solar activity in pre-telescopic times are concentrations of cosmogenic isotopes, such as \element[ ][14]{C} or \element[ ][10]{Be}, in natural archives. From these it is possible to deduce the Sun's open magnetic flux and the sunspot number, as has been demonstrated in numerous publications \citep[e.g.,][]{usoskin2003, usoskin2006a, solanki2004, steinhilber2010}. In fact, when computing the sunspot number from the open flux, the magnetic flux in active regions, ephemeral regions and the total magnetic flux are computed. These are also the ingredients needed to estimate the total and spectral solar irradiance employing the model described by \cite{krivova2007, krivova2010a}. The main hurdle is that in general only the cycle-averaged open flux is known, so that the models originally developed for time-resolved data need to be modified in order to be applicable on these long time scales. For the magnetic flux the relevant relationships were most recently derived by \cite{vieira2010}. Here, we now take the next step and determine the Sun's cycle averaged irradiance.

The aim of this section is to derive the relationship between the cycle-averaged TSI and a solar quantity as closely related to the observations (isotope concentrations) as possible. The open magnetic flux of the Sun is such a quantity. We stress that the open magnetic flux itself produces only a rather minor part of solar irradiance variations, but is closely linked to the quantities which produce larger fractions of the irradiance variations. In particular, the cycle averaged open magnetic flux closely follows the Sun's total magnetic flux.

Before we begin, a brief note on terminology: SATIRE stands for Spectral and Total Irradiance Reconstructions. The basic philosophy of this model family is that all solar irradiance variations on time-scales longer than a day are due to the evolution of the solar surface magnetic field. Members of the SATIRE family of models are SATIRE-S, which models the irradiance during the satellite era with high precision (in its present version since 1974), and SATIRE-T, which computes irradiance during the telescope era (since 1610). Here we introduce the SATIRE-M model, where 'M' stands for millennial time scales.

\subsection{Model of solar irradiance variations}

The Sun's irradiance responds directly to the amount and distribution of magnetic flux on the solar disk, which gives rise to dark features, sunspots, and bright features, faculae and network. In order to quantitatively reproduce solar radiative flux it is sufficient to segment the solar disk, using magnetograms and brightness images, into dark (sunspot umbra and penumbra) and bright (faculae and network) magnetic components as well as the unaffected quiet Sun \citep{krivova2003, wenzler2005, wenzler2006}. The computed irradiance depends both on the area covered by a given type of feature and the spatial location on the solar disk. In the period before magnetograms became available, a simplified version of the model had to be employed. In this model a homogeneous distribution of magnetic features over the solar surface is assumed, so that only the fractional area of the solar disk covered by each of them (global, or disk-averaged filling factor, $\alpha$) and the specific flux, $F$, (i.e. the flux that would be emitted if the whole solar disk was covered by that type of feature) of each feature is required. The evolution of radiative flux at a given wavelengths, $\lambda$, can then be computed as follows \citep{krivova2007,krivova2010a},

\begin{eqnarray}
\label{IMeq01}
F(\lambda, t) & = & \alpha_q(t) F_q(\lambda)  + \alpha_u(t) F_u(\lambda) +  \alpha_p(t) F_p(\lambda)  + \nonumber \\
&  & \alpha_f(t) F_f(\lambda) + \alpha_n(t) F_n(\lambda) \,,
\end{eqnarray}
where the $F_q$, $F_u$, $F_p$, $F_f$, and $F_n$ are the time-independent radiative fluxes of the five component model: quiet Sun (q), sunspot umbra (u), sunspot penumbra (p), faculae (f), and network (n). Following \cite{krivova2007} and others, 
 the spectrum of each component that was calculated by \cite{unruh1999} using the ATLAS9 code of \cite{kurucz1993}. Here, we employ the same model to describe the brightness of facular and network components.

The fraction of the solar disk covered by each component is represented by $\alpha_q$, $\alpha_u$, $\alpha_p$, $\alpha_f$ and $\alpha_n$, respectively. By taking this approach we are assuming a nearly homogeneous distribution of magnetic features on the solar disk.  The filling factor of the quiet Sun can then be written as:

\begin{equation}
\alpha_q(t) = 1 - \alpha_u(t) - \alpha_p(t) - \alpha_f(t) - \alpha_n(t) \,.
\label{IMeq02}
\end{equation}

Substituting Eq. (\ref{IMeq02}) in (\ref{IMeq01}), we obtain

\begin{eqnarray}
F(\lambda, t) &= & (1 - \alpha_u(t) - \alpha_p(t) - \alpha_f(t) - \alpha_n(t)) F_q(\lambda)  +  \alpha_u(t) F_u(\lambda) + \nonumber \\
&  & \alpha_p(t) F_p(\lambda) + \alpha_f F_f(\lambda) +  \alpha_n(t) F_n(\lambda)  \,.
\label{IMeq03}
\end{eqnarray}
This equation can be rewritten as

\begin{eqnarray}
F(\lambda, t) &= & F_q(\lambda)  + \alpha_u(t) \Delta F_u(\lambda) + \alpha_p(t) \Delta F_p(\lambda) +  \nonumber \\
& &\alpha_f(t) \Delta F_f(\lambda) + \alpha_n(t) \Delta F_n(\lambda) \,,
\label{IMeq04}
\end{eqnarray}
where

\begin{equation}
\Delta F_{u,p,f,n}(\lambda) = F_{u,p,f,n}(\lambda) - F_q(\lambda)\,.
\label{IMeq05}
\end{equation}

\subsection{Filling factors for sunspot umbrae and penumbrae}

We compute the filling factors for the umbral and penumbral components by employing the approach described by \cite{krivova2007}. A fixed ratio of 0.2 between umbral and total spot area is assumed following \cite{wenzler2005}, which translates into

\begin{equation}
\alpha_u / \alpha_s = \alpha_u / (\alpha_u + \alpha_p) = 0.2  \,.
\label{IMeq08}
\end{equation}
We compute the sunspot area, i.e. the fraction of the disk covered by all sunspots on the solar disk, by making use of a linear relationship to the sunspot number ($R$) \citep{fligge1997, balmaceda2009, hathaway2010}:

\begin{equation}
\alpha_s =  A_1 R + A_2 \,,
\label{IMeq09}
\end{equation}
where $A_1$ is the proportionality coefficient and $A_2$ is the offset. The key step for extending the irradiance model to the whole Holocene is to compute the sunspot number as well as the magnetic flux in active regions (AR) and ephemeral regions (ER) from the open magnetic flux, since it is the first solar parameter obtained from the chain of models that convert  cosmogenic isotopes data into solar parameters. This problem was first addressed by \cite{usoskin2002} and later updated by \cite{usoskin2003, usoskin2007}, \cite{solanki2004},  and \cite{vieira2010}. In the
 current work we employ the method described by \cite{vieira2010} to compute the total solar surface magnetic flux and the sunspot number from the open flux.   From the information recovered from cosmogenic isotopes, in particular the INTCAL98 \element[ ][14]{C} record \citep{stuiver1998},  we are limited to estimate 10-year averages of the sunspot area and the solar surface magnetic components. Following \cite{vieira2010} we write for decadal values

\begin{equation}
\Rj =a_R \openj + b_R \openjj \,,
\label{IMeq10}
\end{equation}
where $<...>_j$ and $<...>_{j+1}$ are the $j^{th}$ and $j^{th}+1$ 10-year averaged values (or data points) of a variable. Since 10-years is close to the average length of the solar cycle (approximately 11 years for the past 23 cycles),  $j^{th}$ and $j^{th}+1$ also roughly refer to the  $j^{th}$ and $j^{th}+1$ cycles. The coefficients $a_R$ and $b_R$ are given by

\begin{equation}
\label{IMeq11}
a_R = \frac{1}{c \tau_1}
\end{equation}
and

\begin{equation}
\label{IMeq12}
b_R = \frac{1}{c \Delta t} \,,
\end{equation}
where $\Delta t$ is the sampling interval, $\tau_1$ is given by

\begin{equation}
\label{IMeq13}
\frac{1}{\tau_1} = \frac{1}{\tau_{open}^s} - \frac{1}{\Delta t} \,,
\end{equation}
and the constant c is given by

\begin{equation}
\label{IMeq14}
c = \left[ \left( \frac{1}{\tau_{act}^s} + \frac{\tau_{open}^r}{\tau_{open}^s \tau_{act}^r}\right) \tau_{act} + \frac{\tau_{eph}kX}{\tau_{eph}^s} \right] \frac{\epsilon_{act}^{max,21}}{R^{max,21}} \,.
\end{equation}
In Eqs. (\ref{IMeq10}-\ref{IMeq14}), $\tau_{open}^r$ and $\tau_{open}^s$ are the decay times for the rapidly and slowly evolving components of the open flux. $\tau_{act}^s$ and $\tau_{eph}^s$ are the flux transfer times from active and ephemeral regions to the slowly evolving open magnetic flux, while $\tau_{act}^r$ is the flux transfer times from active regions to the rapidly evolving open magnetic flux. $\epsilon_{act}^{max,21}$ and $R^{max,21}$ are the maximum emergence rate of magnetic flux in active regions and the maximum value of the sunspot number observed in cycle 21, respectively. The parameter $k$ is the ratio between sunspot number amplitude and the 10-year averaged value \citep{usoskin2007}. $X$ is the scaling factor for the flux emergence rate in ER relative to active regions. Here, we employ the set of parameter values employed by \cite{krivova2010a} (see Table 1). The sampling interval, $\Delta t$, is 10 years.  The effective decay time scales of active regions (AR), $\tau_{act}$, and ephemeral regions (ER),  $\tau_{eph}$, are given by

\begin{equation}
\frac{1}{\tau_{act}} = \frac{1}{\tau_{act}^0} + \frac{1}{\tau_{act}^s} + \frac{1}{\tau_{act}^r}
\label{IMeq15}
\end{equation}
and
\begin{equation}
\frac{1}{\tau_{eph}} = \frac{1}{\tau_{eph}^0} + \frac{1}{\tau_{eph}^s}  \,,
\label{IMeq15}
\end{equation}
where, $\tau_{act}^0$ and $\tau_{eph}^0$ are the decay time scales of active and ephemeral regions, respectively.

Combining Eqs. (\ref{IMeq08}), (\ref{IMeq09}) and (\ref{IMeq10}), we obtain

\begin{eqnarray}
\alphauj & = &0.2  (A_1  a_R \openj +  A_1 b_R \openjj  + A_2)  \nonumber \\
              & = &a_u \openj + b_u \openjj  + c_u
\label{IMeq15}
\end{eqnarray}
and

\begin{eqnarray}
\alphapj &=& 0.8  (A_1 a_R \openj + A_1 b_R \openjj  + A_2)   \nonumber \\
&=& a_p \openj + b_p \openjj + c_p  \,.
\label{IMeq16}
\end{eqnarray}

The sunspots of a given cycle influence not only the open flux of that particular cycle, but also of the next cycle due to the slow decay of the open flux \citep{solanki2000}. Consequently, it takes the knowledge of the open flux from 2 cycles to get the sunspot areas and the magnetic fluxes of active and ephemeral regions for one cycle, as indicated by Eqs (\ref{IMeq15}) and (\ref{IMeq16}). It means that the part of the magnetic flux that emerges in active and ephemeral regions, and is dragged outward by the solar plasma and permeates the heliosphere, is implicitly represented in Equation (\ref{IMeq10}).  The relative contribution between the $j^{th}$ and $j^{th}+1$ data points (${a_R}/{b_R} = {(\Delta t - \tau_{open}^s)}/{\tau_{open}^s}$) depends on the decay time scale of the slow component of the open flux and on the sampling interval. For the set of parameters employed here (given in Table 1), the contribution of the $j^{th}$ data point relative to that of the $j^{th}+1$ data point is

\begin{equation}
\frac{a_R}{b_R} = \frac{a_u}{b_u} =\frac{a_p}{b_p}  \approx 2.4   \,,
\end{equation}
which suggests that a direct, instantaneous relation between sunspot areas and the open flux is not appropriate.

\subsection{Filling factors for faculae and network}

We compute the 10-year averaged filling factors for faculae ($\alphafj$) and the network ($\alphanj$) from the solar surface magnetic flux employing the same scheme as in previous applications of the SATIRE models \citep{ krivova2003, krivova2007, krivova2010a, wenzler2004, wenzler2005, wenzler2006}. $\alphafj$ and $\alphanj$ are proportional to the magnetic flux in faculae ($\facj$) and network ($\netj$), respectively, until a saturation limit $\phi_{sat}$ is reached. The saturation flux is a free parameter of the SATIRE models, which depends on the resolution and the noise level of the employed magnetic field maps. Observing that the 10-year averaged values $\facj$ and $\netj$ are always smaller than the saturation level, so that they never saturate, we can write

 \begin{equation}
\alphafj = \frac{\facj}{S_{sun} B_{sat,f}} \,,
\label{IMeq17}
\end{equation}
and
 \begin{equation}
\alphanj = \frac{\netj}{S_{sun} B_{sat,n}} \,,
\label{IMeq18}
\end{equation}
where $S_{sun}$ is the Sun's surface area. $B_{sat,f}$ and $B_{sat,n}$ are the saturation field strengths for faculae and network, respectively. Here we follow  \cite{krivova2007} and  \cite{krivova2010a} who incorporated a correction in  $B_{sat,f}$ that takes into account that the faculae mainly appear in the active belts.

The magnetic flux in active regions ($\actj$) is the sum of the flux in sunspots (umbrae plus penumbrae) and faculae. Thus, the magnetic flux in faculae is given by

\begin{equation}
\facj = \actj - \uj - \pj \,.
\label{IMeq19}
\end{equation}

Following \cite{krivova2007}, we compute

\begin{equation}
\uj + \pj=   \alphauj  [B_{z,u}] S_{sun}+  \alphapj [B_{z,p}] S_{sun}\,,
\label{IMeq20}
\end{equation}
where $ [B_{z,u}]$ and $ [B_{z,p}]$ are the area-averaged vertical components of the magnetic field strength in umbrae and penumbrae, respectively. Note that since $ \alphauj$  and  $\alphapj$ are indirectly obtained from $<R>_j$, it is assumed implicitely that averaged over 10 years there is no major difference between the number of {sunspots} on the visible part of the solar surface (solar disk) and on the hidden part.

From Eq. (25) of \cite*{vieira2010}, the magnetic flux in active regions is approximately given by

\begin{equation}
\actj \approx \tau_{act} \frac{\epsilon_{act}^{max,21}}{R^{max,21}} \left< R \right>_j \,.
 \label{IMeq21}
\end{equation}

Combining Eqs. (\ref{IMeq15}),  (\ref{IMeq17}), (\ref{IMeq18}),  (\ref{IMeq19}), (\ref{IMeq20}), and (\ref{IMeq21}), we obtain

\begin{equation}
\alphafj =a_f \openj + b_f \openjj + c_f\,.
\label{IMeq23}
\end{equation}

The sum of ephemeral regions flux and of the open flux describes the evolution of the network:

\begin{equation}
\netj = \ephj + \openj \,.
\label{IMeq24}
\end{equation}
From Eq. (29) of \cite*{vieira2010}, the 10-year averaged magnetic flux in ephemeral regions is

\begin{eqnarray}
\label{eq33}
\ephj &\approx& \tau_{eph} \frac{\epsilon_{act}^
{max,21}}{R^{max,21}} k X \Rj  \nonumber \\
 &\approx& \tau_{eph} \frac{\epsilon_{act}^
{max,21}}{R^{max,21}} k X \left( a_R \openj + b_R \openjj \right) \,.
\label{IMeq25}
\end{eqnarray}
The contribution from the next i.e. $(j+1)$-st cycle is due to the extended period over which ephemeral regions emerge. From Eqs. (\ref{IMeq24}) and (\ref{IMeq25}),

\begin{equation}
\netj =  a_{\phi_n} \openj + b_{\phi_n} \openjj \,,
\label{IMeq26}
\end{equation}
where

\begin{equation}
a_{\phi_n} =  \tau_{eph} \frac{\epsilon_{act}^{max,21}}{R^{max,21}} k X a_R + 1
\label{IMeq27}
\end{equation}
and
\begin{equation}
b_{\phi_n} =  \tau_{eph} \frac{\epsilon_{act}^{max,21}}{R^{max,21}} k X b_R  \,.
\label{IMeq28}
\end{equation}
Substituting Eq. (\ref{IMeq26}) into (\ref{IMeq18}), we obtain

\begin{eqnarray}
\alphanj &=& \frac{a_{\phi_n} \openj + b_{\phi_n} \openjj}{S_{sun} B_{sat,n}} \nonumber \\
&=& a_{n} \openj + b_{n} \openjj \,.
\label{IMeq29}
\end{eqnarray}
While the relative contributions between $j^{th}$ and $j^{th} + 1$ data points of the open flux for the umbral, penumbral, and facular filling factors depend on the decay time of slowly evolving open flux and the sampling time, the relative contribution to the network is given by

\begin{equation}
\frac{a_n}{b_n}=  \frac{\Delta t}{\tau_{open}^s}  +    \frac{R^{max,21} c \Delta t }  { \epsilon_{act}^{max,21} \tau_{eph} k X}   - 1    \,.
\label{IMeq30}
\end{equation}
For the set of parameters employed here, this ratio is approximately 3.3. This indicates that although the contribution from the next cycle is lower than that from the cycle for which we want to know $\alphanj$, it is still necessary to know the next state in order to evaluate the filling factor of the network accurately.

\subsection{Estimate of the total solar irradiance}

The solar irradiance at wavelength $\lambda$ can be computed by substituting the expressions for the filling factors obtained for the components into Eq. (\ref{IMeq04})

\begin{equation}
\Fj  =  a_{F} (\lambda) \openj+ b_{F}(\lambda)  \openjj + F_q(\lambda)  \,,
\label{IMeq31}
\end{equation}
where

\begin{equation}
a_{F}(\lambda) = a_u \Delta F_u(\lambda) + a_p \Delta F_p(\lambda) + a_f \Delta F_f(\lambda) + a_n \Delta F_n(\lambda) \,,
\label{IMeq32}
\end{equation}
and

\begin{equation}
b_{F}(\lambda) = b_u \Delta F_u(\lambda)  + b_p \Delta F_p(\lambda)  + b_f \Delta F_f(\lambda)  + b_n \Delta F_n(\lambda)  \,.
\label{IMeq33}
\end{equation}

In order to obtain the total solar irradiance one can either integrate $\Fj$ over all wavelengths or right from the beginning work with wavelength integrated radiative fluxes and flux differences

\begin{equation}
F_{q}^t = \int_{0}^{\infty}  F_{q}(\lambda) d\lambda \,,
\label{IMeq34}
\end{equation}
and

\begin{equation}
\Delta F_{u,p,f,n}^t = \int_{0}^{\infty} \left[ F_{u,p,f,n}(\lambda) - F_{q}(\lambda) \right] d\lambda \,.
\label{IMeq34}
\end{equation}

Replacing the $F_q$ and $\Delta F_{i}$ in Eqs. (\ref{IMeq31})-(\ref{IMeq33}) by $F_q^t$ and $\Delta F_{i}^t$, $i = u,p,f,n$, one then obtains the total solar irradiance averaged over 10 years,  $<F^t>_j$. We refer for clarity to  $<F^t>_j$ and  $<F_i^t>_j$, $i = u,p,f,n$, as $TSI$ and $TSI_i$.

Hence, the solar spectral and total irradiance can be computed from the $j^{th}$ and $j^{th}+1$ values of the open flux in a relatively straightforward manner.

\subsection{Reconstruction of the open flux from cosmogenic data}
\label{Sec:cosmogenic}
{ 
The open flux in the past can be reconstructed from data of cosmogenic isotopes measured in terrestrial archives. Here we use data from \element[ ][14]{C}measured in tree-rings for the past millennia as compiled by the INTCAL04 collaboration \citep{reimer04}. We have applied a commonly used approach of inverting the cosmogenic isotope data into cosmic ray flux. This approach, proposed \citet{usoskin2003,usoskin2004}, is now a standard method used to reconstruct of the solar or heliospheric parameters in the past \citep[e.g.][]{solanki2004,usoskin2006a,usoskin2007,vonmoos2006,muscheler2007, steinhilber2008}. The method includes the following steps.
\begin{enumerate}
\item
First, the measured values of $\Delta$\element[ ][14]{C} are converted, using a carbon cycle model, into the global $^{14}$C production rate, Q$_{^{14}{\rm C}}$.
We used a standard multi-box carbon cycle model \citep{siegenthaler1980} parameterized using the approach of \citet{usoskinkromer2005}.

\item
Considering prescribed geomagnetic field data and a model of \element[ ][14]{C} production in the atmosphere \citep{castagnoli1980}, the
 production rate Q$^{14}$C was converted into the cosmic ray spectrum quantified via the modulation potential $\Phi$.

\item
Using a standard theory of cosmic ray transport in the Heliosphere one can relate the modulation potential $\Phi$ to the open magnetic flux assuming
 a constant 11-yr averaged solar wind speed \citep[e.g.][]{usoskin2002}.
\end{enumerate}
Here we give a brief description of the method, while full details are provided elsewhere \citep{solanki2004,usoskin2007}.

The uncertainties of the reconstructed open flux $\sigma_{\phi_{\rm open}}$ have several sources.
The major source of the systematic error is related to the uncertainties of the $^{14}$C production rate
 in the pre-Holocene period.
Other sources of systematic uncertainties are related to the paleomagnetic models (considered below in detail),
 uncertainties of the carbon cycle model, uncertainties of the $^{14}$C production model, and
 uncertainties of the cosmic ray heliospheric transport model.
Random errors are related to measurement errors in $\Delta^{14}$C and in geomagnetic data.
All these uncertainties are described in detail by \citet[][Supplementary matarial]{solanki2004}.
The total uncertainty in the open flux $\sigma_{\phi_{\rm open}}$ is of the order of $0.2\cdot 10^{14}$ Wb for the last five millennia with relatively good geomagnetic data and rises up to $0.5\cdot 10^{14}$ Wb in the early Holocene.

We note that computation of the open flux from $^{14}$C is not a part of the present work as it was performed using the same
 algorithms and codes as in \cite{usoskin2007}, but now applying different paleomagnetic models. { We point out that the 14C-based open flux is consistent with the {\bf \element[ ][44]{Ti}} data from meteorites \citep{usoskin2006ti}}.
}

\section{Results}

\subsection{Model setup and validation}

The equations modeling the evolution of TSI during the Holocene have been derived in the previous section. In this section, we present some tests. We compute the coefficients for the solar spectral and total irradiance model for the Holocene according to the parameters of the SATIRE-T model presented by \cite{krivova2010a}. Table \ref{table2} lists the coefficients entering the equations given in Sect. 2 that are used to compute the solar irradiance from the open magnetic flux (which is deduced from \element[ ][14]{C} data). We stress that we have not introduced new free parameters when simplifying the model to work with 10-year averaged values. Thus, uncertainties in the parameters are a heritage from the SATIRE-T model described by \cite{krivova2010a}.

\begin{figure*}
\centering
\includegraphics[width=13cm,clip=true, viewport=0.5cm 10cm 18.5cm 27cm]{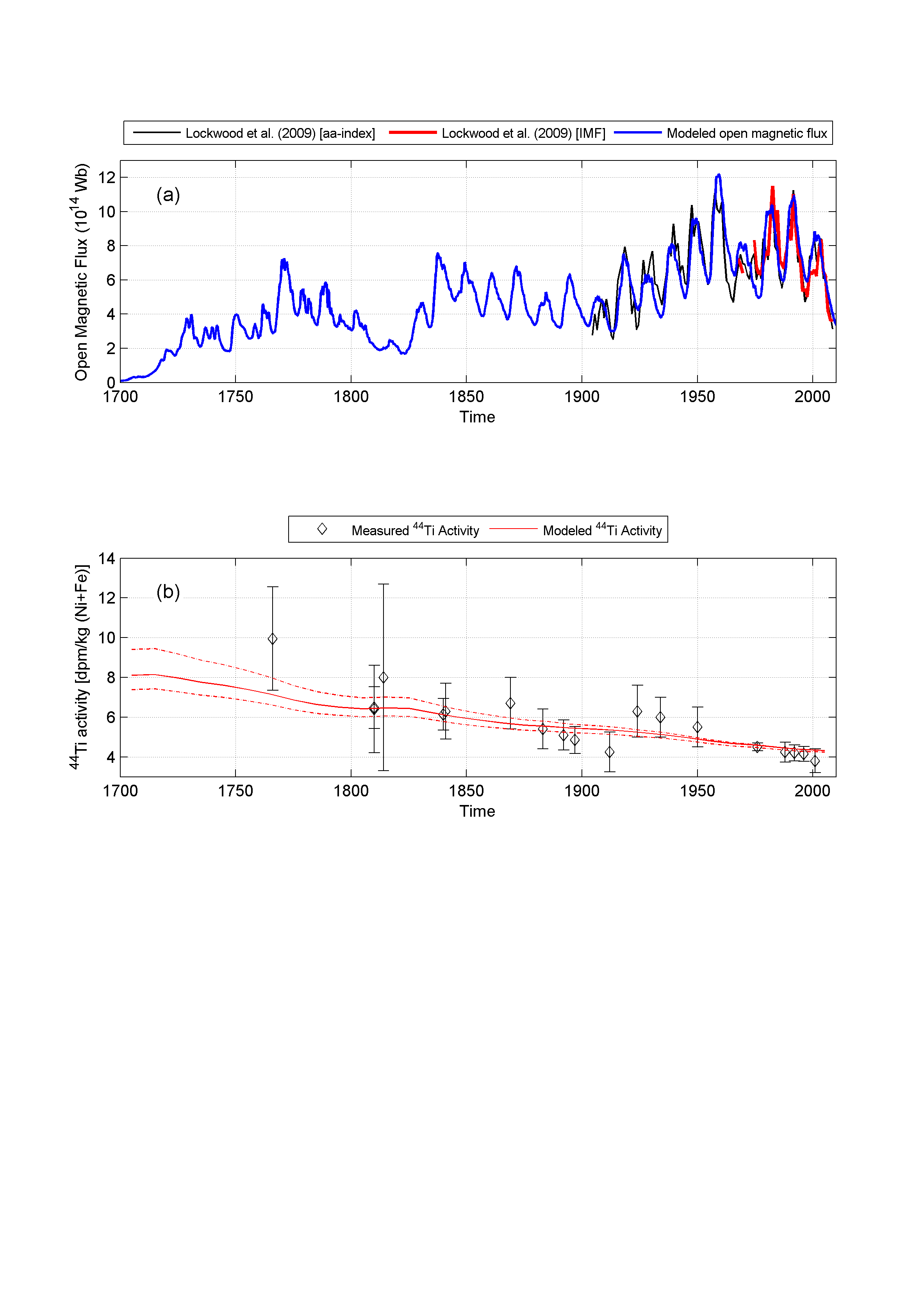}
\caption{(a) Evolution of the solar open magnetic flux. The black and red lines present the estimates of the open flux from the geomagentic aa-index and from observations of the interplanetary magnetic field \citep{lockwood2009a,lockwood2009b}, respectively. The blue line displays the modeled open flux from SATIRE-T. (b) \element[][44]{Ti} activity in stony meteorites as a function of the time of fall \citep[see][and references therein]{tarico2006}. Error bars refer to 1-sigma uncertainties. The red line corresponds to the modeled  \element[][44]{Ti} based on the SATIRE-T open flux according to \cite{usoskin2006ti}.  The \element[ ][44]{Ti} activity is given in disintegrations per minute per kilogram (dpm kg$^{-1}$) of \element[][]{Ni} $+$ \element[][]{Fe}. The dashed red lines correspond to the estimated 1-sigma uncertanty of the model}.
\label{figure_ti_fo}
\end{figure*}

\cite{krivova2010a} constrained the parameters of the model by comparing the model output with observations and reconstructions of five solar variables. While two data sets are magnetic flux variables, three data sets are related directly to the solar irradiance. The total solar surface magnetic flux was constrained by comparing the model output with estimates carried out at the Mt. Wilson Solar Observatory (MWO), National Solar Observatory Kitt Peak (KP NSO) and Wilcox Solar Observatory (WSO) over cycles 20 to 23 \citep{arge2002, wang2005}. The open flux was compared to the reconstruction based on the geomagnetic aa-index \citep{lockwood2009a, lockwood2009b} from 1904 to 2008. These two data sets provided constraints on the amplitude of the magnetic activity cycle and its long-term evolution. We stress that the evolution of the open flux as reconstructed based on the aa-index, which nearly doubled during the last century, is well reproduced by the magnetic field model (see Figure \ref{figure_ti_fo}a). In addition to the reconstruction based on the geomagnetic aa-index, the long-term evolution of solar magnetic activity can also be constrained by cosmogenic isotopes deposited in terrestrial and extra-terrestrial archives \citep{solanki2004, usoskin2006ti}. While terrestrial archives can span the entire Holocene, the processes leading to the deposition of the cosmogenic isotopes are affected by the evolution of, e.g., the Earth's magnetic field and of the highly coupled atmosphere-ocean system. Extraterrestrial archives, such as meteorites, are not affected by such processes and can be employed to test the long-term trend observed in the modeled open flux. In order to substantiate the analysis carried out by \cite{krivova2010a} and \cite{vieira2010}, we test the reconstruction of the open flux by comparing the activity of cosmogenic isotope \element[ ][44]{Ti} in meteorites, which fell during the past 240 years, and the modeled activity based on the open flux computed in the framework of the above models. {We carried out the analysis using the same method as 
described by Usoskin et al. (2006b) including evaluation of all known sources of uncertainties.} Figure \ref{figure_ti_fo}b presents a comparison between the \element[ ][44]{Ti} measurements in 19 meteorites, indicated by diamonds, and the modeled \element[ ][44]{Ti} activity at the time of the meteorite's fall (see details in \cite{usoskin2006ti} and \cite{tarico2006}). The \element[ ][44]{Ti} activity is given in disintegrations per minute per kilogram (dpm kg$^{-1}$). The $\chi^2$ is about 7.6 for the 19 data points, i.e. about 0.4 per degree of freedom, that is comparable to the results previously discussed by \cite{usoskin2006ti}. Thus, the open flux modeled employing the parameters of the SATIRE-T model follows the data reasonably well, which leads us to conclude that the parameterization used is consistent with the available longer term solar data.

\begin{figure*}
\centering
\includegraphics[width=13cm,clip=true, viewport=0.5cm 2cm 18.5cm 27cm]{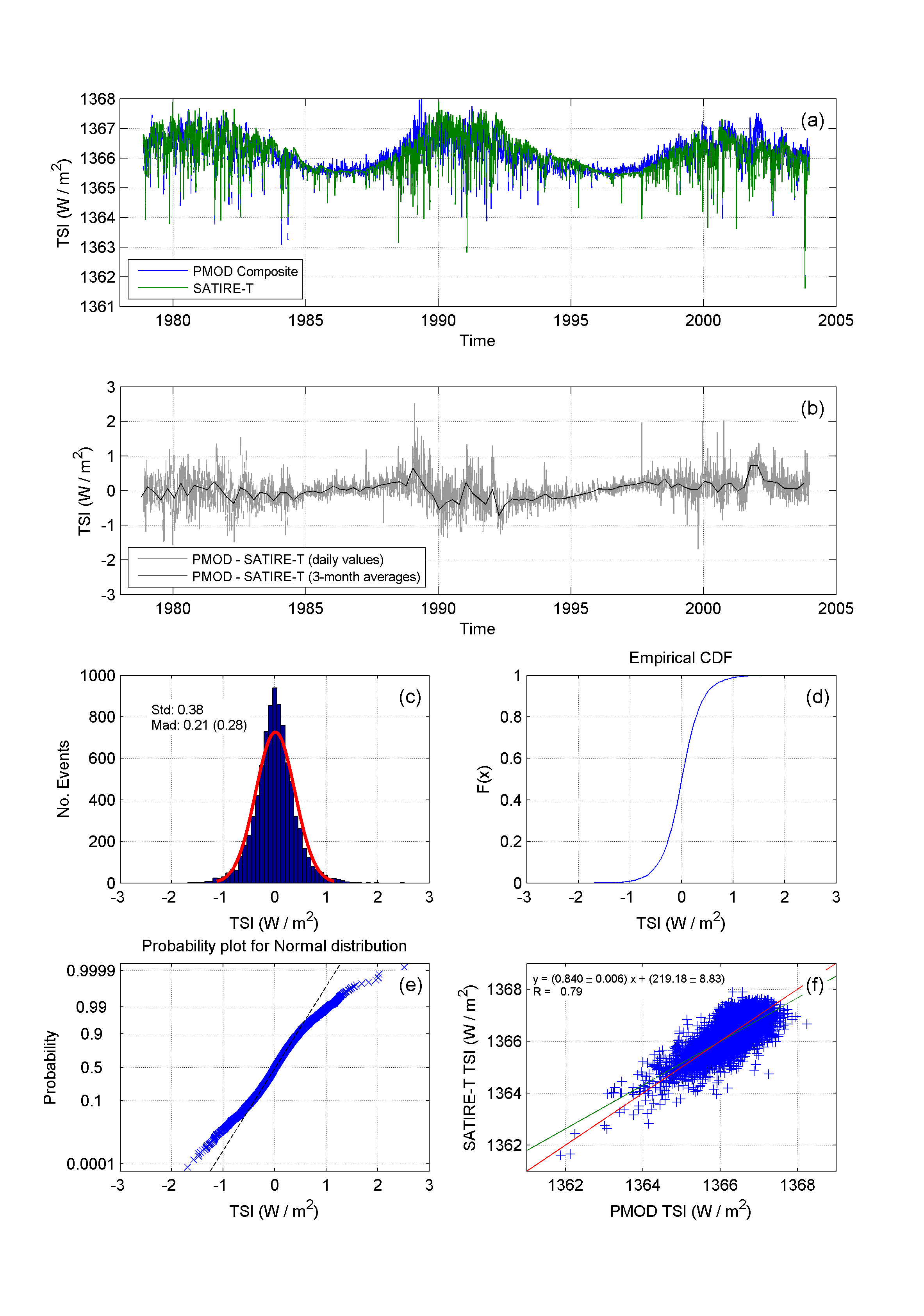}
\caption{(a) Comparison between the observed (PMOD Composite; blue line) and the modeled (SATIRE-T; green line) TSI from Nov/1978 to Dec/2003. (b) Difference between the observed and modeled TSI. The gray and black lines present the daily and 90-day averaged values, respectively. (c) Distribution of the difference between daily values of the observed and modeled TSI (model error). The standard deviation and the median absolute deviation are indicated in the panel. The red curve represents the corresponding probability distribution function for the normal distribution with the mean and standard deviation of the model error. (d) Cumulative distribution function (CDF) for the difference between observed and modeled TSI. (e) Probability plot for normal distribution with the mean and standard deviation of the model error. (f) Scatter plot of the observed versus modeled TSI. The red line indicates the set of expectation values of the model. The correlation coefficient and the coefficients of the best linear fitting (green line) are presented in the panel. The p-value for testing the hypothesis of no correlation is lower than 0.5, which indicates that the correlation is significant. }
\label{hol_error_tsi}
\end{figure*}

The model discussed by \cite{krivova2010a} was further constrained by requiring the computed TSI variations to match the PMOD composite \citep{Frohlich2006}, which is obtained from space-based observations since 1978. Here, we employ (among other measurements) the difference between the SATIRE-T model output and the PMOD composite to estimate the uncertainty of the TSI model for the Holocene. We refer to this difference as the model error. Figure \ref{hol_error_tsi}a presents a comparison between the observed (PMOD composite; blue line) and the modeled (SATIRE-T; green line) TSI. The plot covers the period from Nov./1978 to Dec./2003, which was employed by \cite{krivova2010a} to constrain the SATIRE-T model. Note that as the absolute value of the TSI is not well determined we offset the model to match the PMOD composite level, although observations employing the SORCE/TIM instrument suggest that the absolute value during the minimum between cycles 23-24 is approximately 4-5 W/m$^2$ lower than that measured by other current instruments {\citep{kopp2005, kopp2011}}. Figure \ref{hol_error_tsi}b presents the difference between the PMOD composite and the adjusted SATIRE-T model. The gray and black lines display daily values and 90-day means, respectively. Figure \ref{hol_error_tsi}c presents the distribution of the model error. The standard deviation ($\sigma$) and the median absolute deviation (MAD) are indicated in the panel. The MAD value is defined as $median(abs(X-median(X)))$, where $X$ is the sample population. The red line represents the corresponding normal distribution for the estimated mean and standard deviation. While the empirical cumulative distribution function (CDF) of the model error is presented in Figure \ref{hol_error_tsi}d, the probability plot for a Normal Distribution is presented in Figure \ref{hol_error_tsi}e. In the probability plot, the marks ('x') indicate the empirical probability versus the model error values. Note that the scale of the ordinate axis is not linear. The distance between the ordinate tick marks is proportional to the distance between the quantiles of a normal distribution. The dashed line connects the $25^{th}$ and the $75^{th}$ quantiles. The assumption of normality is reasonable if the data points are near the dashed line. However, this plot clearly indicates that the distribution is not normal. This is a consequence of the procedure employed to constrain the model, which reduced the weight of outliers to compute the $\chi^2$. Here, we use the MAD, which is a more robust estimator than the standard deviation and variance, as a measure of the dispersion of the model error distribution. In order to estimate the dispersion from the MAD value we multiply the MAD value by the scale parameter for the normal distribution \citep{mosteller1977}. The value obtained is approximately 0.28 $W/m^2$. Note that this uncertainty applies to daily data, whereas we are dealing with decadal averages. For these the difference between reconstruction and observations is very much smaller. Nonetheless, we retain 0.28 $W/m^2$ as the model error, since it corresponds roughly to the uncertainty in the Maunder Minimum level \citep{krivova2007}. Figure \ref{hol_error_tsi}f presents the scatter plot of the observed (PMOD Composite) versus modeled TSI (SATIRE-T). The green line presents the best linear fit to the data set. The coefficients and their uncertainties are indicated in the figure as well as the correlation coefficient (Rc = 0.79).

We validate the SATIRE-M model during the telescope era by comparing its output with that of the SATIRE-T model. The two models are not independent, so that with these comparisons we mainly test the simplifications introduced in order to be able to adequately deal with the 10-year averaged data. To compare both models, first we compute 10-year averages from 1700 AD to 2000 AD of the open flux, total magnetic flux and TSI as well as of the contributions from faculae, network and sunspots (umbrae + penumbrae) to TSI from the SATIRE-T model. We then employ the 10-year averaged open flux from SATIRE-T as input to SATIRE-M. This allows us to compute the contribution of each atmospheric component to TSI employing SATIRE-M.

\begin{figure*}
\centering
\includegraphics[width=13cm,clip=true, viewport=0.5cm 2cm 18.5cm 27cm]{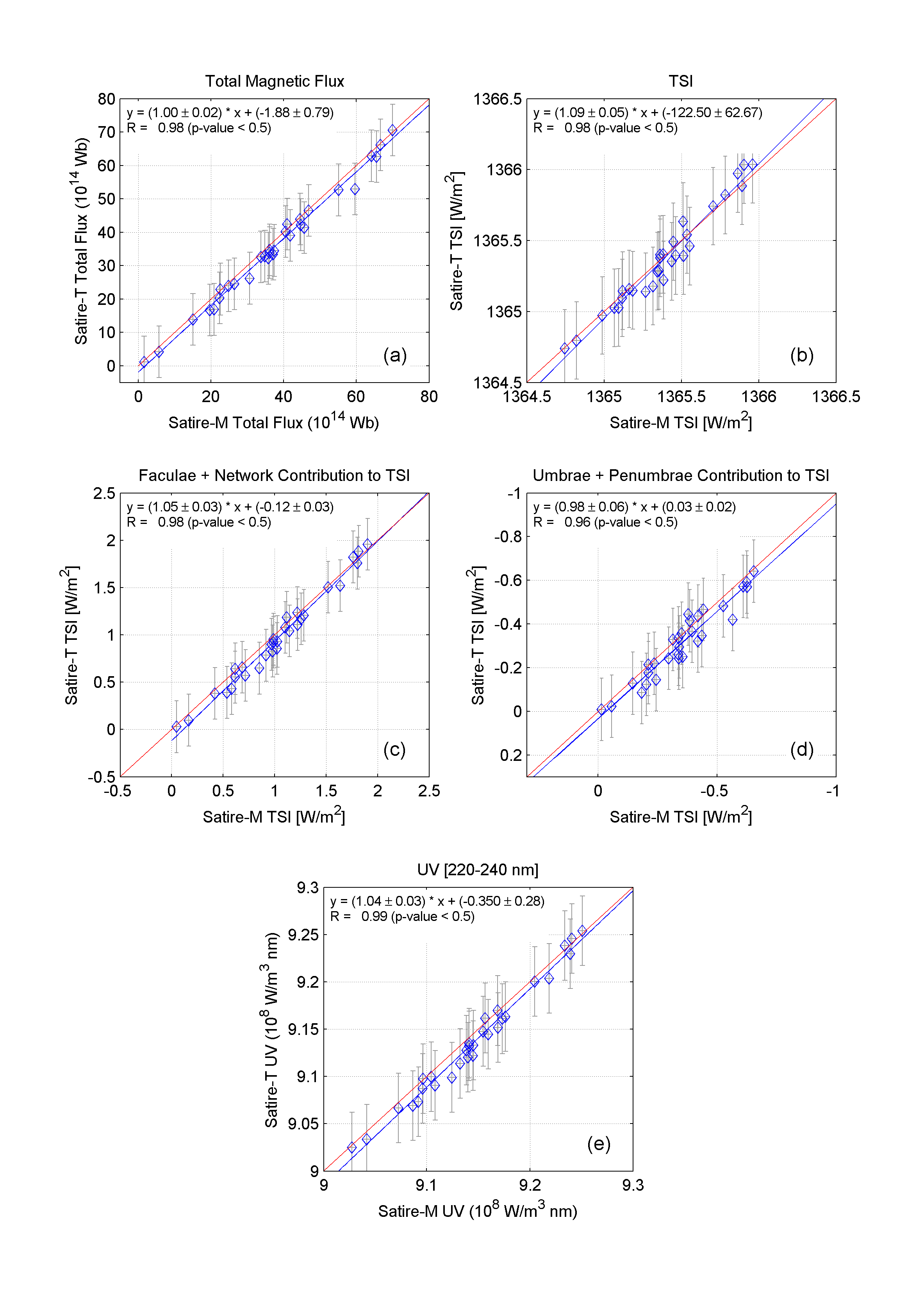}
\caption{Comparison between 10-year averaged values of the parameters computed employing the SATIRE-M and SATIRE-T models from 1700 AD to 2000 AD. The following scatter plots are presented: (a) Total magnetic flux; (b) Total solar irradiance; (c) contribution of the bright components  (faculae + network) to the total solar irradiance; (d) Sunspot (umbra + penumbra) contribution to TSI; (e) UV integrated in the range 220-240 nm. The red lines present the set of expectation values for the models and the blue lines are the best linear fits. The correlation coefficient and the coefficients of the linear regression are presented in the panels.}
\label{holfig1}
\end{figure*}

Figure \ref{holfig1} presents a comparison between the output of the models SATIRE-T and SATIRE-M. Each panel contains the following elements: The red lines indicate the set of expectation values for the model, while the blue lines are the best fit lines. The linear model coefficients and the correlation coefficient are also shown in each panel. The vertical error bars correspond to the dispersion of the SATIRE-T model error computed from the MAD value of the SATIRE-T error distribution and corrected for the normal distribution. The estimate of the SATIRE-M error is discussed in the next subsection. In addition to the dispersion of the TSI error distribution, the other constraints imposed by \cite{krivova2010a} can also be employed to compute the dispersion of the model error distributions of the total magnetic flux and components of the TSI. Additionally to the constraint imposed by the PMOD composite, \cite{krivova2010a} constrained the amplitude of the facular contribution to TSI by comparing the model output with the estimate computed by \cite{wenzler2006}. The dispersion of the difference between the models is about 0.27 W/m$^2$ when corrected for the normal distribution. Although the sunspot deficit computed by \cite{wenzler2006} was not employed to constrain the SATIRE-T model, it could be employed to estimate the uncertainty of the sunspot (umbral + penumbral) contribution to TSI. In this case, the dispersion is about 0.14 W/m$^2$. In order to ensure that not only the total irradiance is reproduced correctly, we also compare the solar irradiance integrated over the wavelength range 220-240 nm \cite[see][]{krivova2009a,krivova2010b} with the SATIRE-T model output by \cite{krivova2010a}.

The SATIRE-M model reproduces quite accurately the evolution of the total magnetic flux as computed by the SATIRE-T model ($Rc = 0.98$; $slope = 1.00 \pm 0.02$; Figure \ref{holfig1}), but it slightly overestimates the magnitude of the total magnetic flux ($offset = -1.88 \pm 0.79$ ($\times 10^{14}$ Wb)). The SATIRE-M model also reproduces well the total solar irradiance computed employing SATIRE-T ($Rc = 0.98$), but slightly overestimates low values and underestimates high values, which result in a $slope = 1.09 \pm 0.05$). A similar behavior is also observed when comparing the facular and network brightness excess. The coefficient of correlation is approximately 0.98 while the $slope$ is approximately $1.05$. The sunspot (umbrae + penumbrae) brightness deficit is also well reproduced but the coefficient of correlation ($Rc = 0.96$) is lower than the one obtained for the comparison of the brightness excess from faculae and network. The $slope$ of the linear regression is close to unity ($slope = 0.98$). The UV irradiance integrated from 220 to 240 nm follows the behavior of the facular and network contribution of the TSI, with a high coefficient of correlation ($Rc = 0.99$).

The correlation coefficients for all cases exceed 0.96 and the differences between the models display nearly Gaussian distributions. Thus, we conclude that the 10-year averages of the $TSI$ (and UV irradiance) can be obtained from a linear combination of the $j^{th}$ and $j^{th}+1$ observations of the open magnetic flux with reasonable accuracy.


\begin{table*}
\caption{Parameters of the TSI model presented by \cite{krivova2010a} and employed here.}              
\label{table1}      
\centering                                      
\begin{tabular}{l c r c}          
\hline\hline
Parameter	& Symbol & Value  & Unit\\
\hline
AR Flux decay time scale	               		& $\tau_{act}^0$  	& 0.30     	&  year \\
AR Flux to Slow Open Flux transfer time scale  	& $\tau_{act}^s$	& 71.2       	&  year \\
AR Flux to Rapid Open Flux transfer time scale 	& $\tau_{act}^r$       & 2.1   		&  year\\
ER  Flux decay time scale	               		& $\tau_{eph}^0$      & 0.0016        	& year\\
ER Flux to Slow Open Flux transfer time scale  	& $\tau_{eph}^s$      & 17.8   	& year\\
Rapid Open Flux decay time scale	       		& $\tau_{open}^r$    & 0.16 		& year\\
Slow Open Flux decay time scale	               	& $\tau_{open}^s$   & 2.97   	& year\\
ER amplitude factor	                       			& $X$	     	           & 78 		& 	\\
ER cycle extension parameter	               		& $c_x$	           & 5.01      	& 	\\
\hline
Saturation flux in faculae				& $\phi_{sat,f}$	& 260.2	& $G$\\
Saturation flux in network				& $\phi_{sat,n}$	& 800 		& $G$\\
\hline
Umbral  brightness deficit					& $\Delta I_u$ 	&   -871.8	& $W m^{-2}$ \\
Penumbral  brightness 	deficit				& $\Delta I_p$ 	&-277.1 	& $W m^{-2}$ \\
Facular   brightness excess					& $\Delta I_f$ 		&   85.6	& $W m^{-2}$ \\
Network brightness  excess					& $\Delta I_n$ 	&   85.6	& $W m^{-2} $ \\
\hline
\end{tabular}
\end{table*}


\begin{table*}
\caption{Coefficients of the model to compute TSI for the Holocene (SATIRE-M). The dependent variables (listed under the heading 'Quantity') are computed from the equation $x = a  \openj + b  \openjj + c$ }
\label{table2}
\centering
 \begin{tabular}{l r c r c  r c r r}
\hline\hline
Quantity	& $a$  & & $b$ &  & $c$ & & $a/b$	\\
\hline
$\phi_{act}$ &      2.4 & &        1 & & & &     2.37 \\
$\phi_{eph}$ &      2.5 & &      1.1 & & & &     2.37 \\
$\alpha_u$ & 2.79e-005 & [$Wb^{-1}$] & 1.17e-005 & [$Wb^{-1}$] & -9.3e-007 & &     2.37 \\
$\alpha_p$ & 1.11e-004 & [$Wb^{-1}$] & 4.70e-005 & [$Wb^{-1}$] & -3.7e-006 & &     2.37 \\
$\alpha_f$ & 1.99e-003 & [$Wb^{-1}$] & 8.39e-004 & [$Wb^{-1}$] & 2.6e-005 & &     2.37 \\
$\alpha_n$ & 7.18e-004 & [$Wb^{-1}$] & 2.16e-004 & [$Wb^{-1}$] & & &     3.32 \\
$ TSI_u$ &   -0.024  & [$W m^{-2} Wb^{-1}$] &    -0.01  & [$W m^{-2} Wb^{-1}$] & 8.14e-004 & [$W m^{-2}$] &     2.37 \\
$TSI_p$ &   -0.031  & [$W m^{-2} Wb^{-1}$] &   -0.013  & [$W m^{-2} Wb^{-1}$] & 1.04e-003 & [$W m^{-2}$] &     2.37 \\
$TSI_f$ &      0.1  & [$W m^{-2} Wb^{-1}$] &    0.043  & [$W m^{-2} Wb^{-1}$] & 1.37e-003 & [$W m^{-2}$] &     2.37 \\
$TSI_n$ &    0.061  & [$W m^{-2} Wb^{-1}$] &    0.019  & [$W m^{-2} Wb^{-1}$] & &  &     3.32 \\
$TSI$ &     0.11  & [$W m^{-2} Wb^{-1}$] &    0.039  & [$W m^{-2} Wb^{-1}$] &  1364.72 & [$W m^{-2}$] &     2.83 \\
$UV${[220-240 nm]} &   2e+006  & [$W m^{-3} nm Wb^{-1}$] & 7.4e+005 & [$W m^{-3} nm Wb^{-1}$] &   9e+008 & [$W m^{-3} nm$] &     2.67 \\
\hline
\end{tabular}
\end{table*}

\subsection{Reconstructions of the solar irradiance}
\label{Sec:geom}
\begin{figure*}
\centering
\includegraphics[width=13cm,clip=true, viewport=0.5cm 2cm 20cm 27cm]{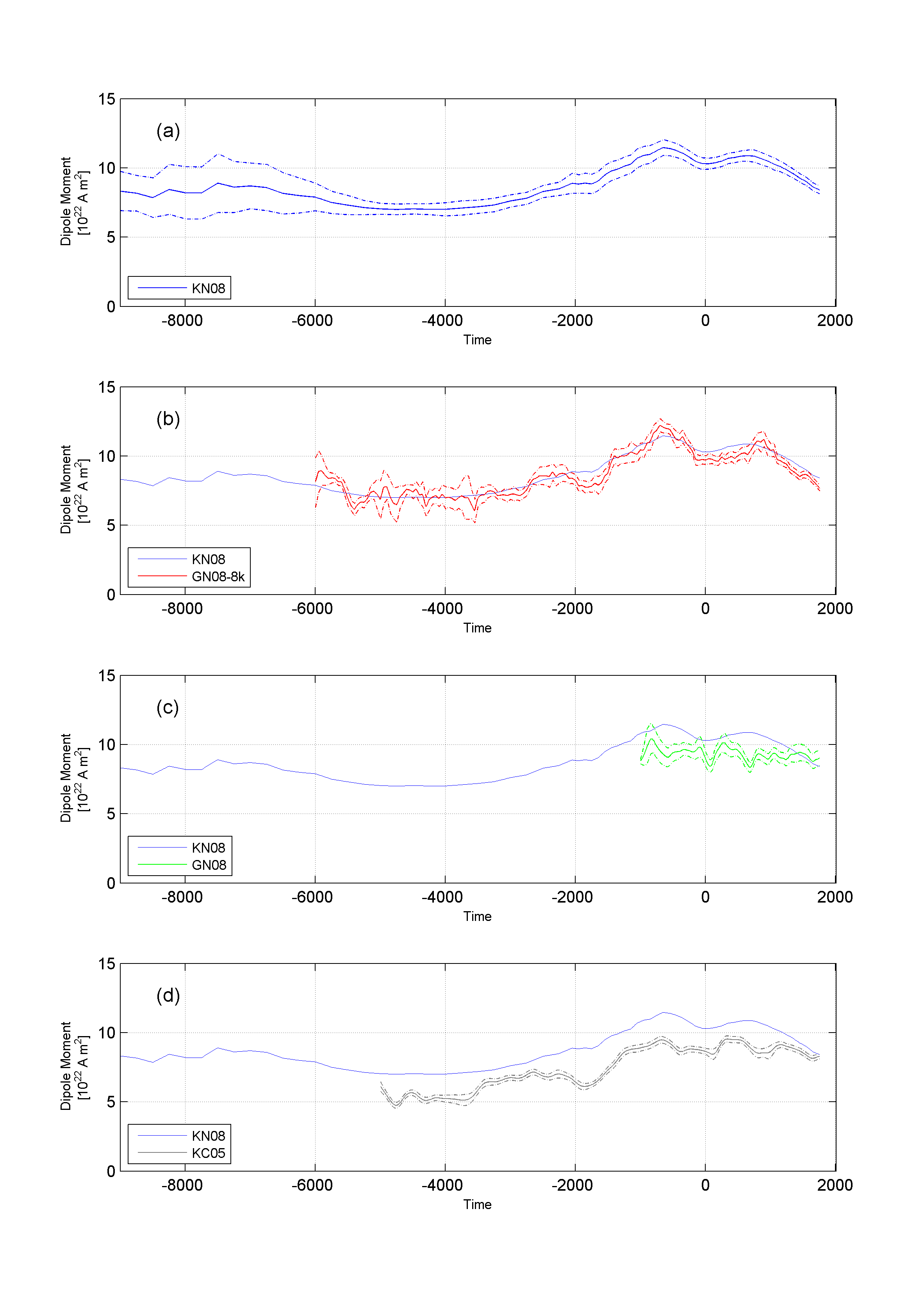}
\caption{Reconstructions of the geomagnetic field intensity. The VADM reconstructions by \cite{knudsen2008} and \cite{genevey2008} are presented in panels (a) and (b). The VDM reconstructions by \cite{genevey2008} and \cite{korte2005} are shown in panels (c) and (d), respectively. The dashed lines represent 1-$\sigma$ statistical errors. For reference, the VADM reconstruction by \cite{knudsen2008} is presented in all panels.}
\label{FigMagLong}
\end{figure*}

As shown in Sect. 2, an estimate of the evolution of the TSI during the Holocene can be obtained from a linear
combination of the $j^{th}$ and $j^{th} + 1$ decadal values of the open flux, which can be computed from cosmogenic
isotopes recorded in natural archives.
{ 
As described in Sect. ~\ref{Sec:cosmogenic}, the reconstructions of the open magnetic flux used here are based  on the INTCAL04 $\Delta$\element[ ][14]{C} data set 
\citep{reimer04} employing almost exactly the same algorithm as the one used by \cite{solanki2004} and
\cite{usoskin2007,usoskin2009}.
}
In addition to the uncertainties in computing the TSI from decadal values of the open flux employing SATIRE-T, there are also
 uncertainties related to the computation of the open flux from cosmogenic isotopes { (see Sect.~\ref{Sec:cosmogenic})}. 
The largest uncertainty is posed by the evolution of the geomagnetic field \citep{usoskin2006a, usoskin2007, snowball2007}. 
Information of the geomagnetic field prior to the era of direct measurements can be retrieved from archeological artifacts,
 lava flows, and sediments \citep[see ][and references therein]{donadini2010}. 
In order to reconstruct the evolution of the geomagnetic field it is necessary to determine at least one component of the
 magnetic field and the period that the information was stored in the natural archive. 
Both parameters introduce uncertainties to the determination of the Earth's magnetic field, as discussed by \cite{donadini2010}. 
Paleomagnetic reconstructions of the Earth's dipole moment can be based on different assumptions. 
{ 
Precise determination of the true dipole moment is hardly possible before the era of systematic geomagnetic
 measurements started in the 19th century. 
When the quality and spread of data is still good enough, a VDM (virtual dipole moment) can be estimated, 
 assuming that the geomagnetic field can be approximated by a tilted geocentric dipole. 
When the spatial coverage of paleomagnetic samples becomes too patchy, only the VADM (virtual axial dipole moment) can be estimated
 by assuming that the geomagnetic axis is aligned with the geographical axis. 
Both VDM and VADM form a proxy for the true {dipole moment (DM)}, but they include also non-dipole contributions assigned to the dipole strength. 
Since the  VADM assumes the dipole to be aligned with the geographical axis, its values can be strongly distorted if the dipole axis is tilted. 
VADM and VDM estimates may differ significantly from the true DM, especially on short time scales due to the non-dipole influence
 in the VADMs caused by a strong regional bias in data distribution \citep{korte2005}.
Generally VDM models give a better approximation to the true DM when the data coverage is sufficient
 but tend to underestimate it with spatially limited data sets. 
In contrast, VADM models tend to overestimate the true moment, especially for insufficient data coverage and tilted dipole axis \citep{donadini2010}.
Therefore, we apply the following approach to estimate the systematic uncertainty of the results due to paleomagnetic reconstructions: we consider the VDM to be a lower bound and VADM an upper bound to the true dipole moment. In order to be able to provide a single reconstruction of TSI over a certain period of time, we simply average the output from VDM and VADM based reconstructions of TSI. We do not claim that the average of the two groups is the most probable value, although  the most probable value lies between these bounds. Table 3 presents the geomagnetic models employed here, references, and the time coverage. {Note that we have employed the global reconstructions from \cite{genevey2008}}. Figure \ref{FigMagLong} displays the reconstructions of the geomagnetic field intensity, including  VADM reconstructions by \cite{knudsen2008} and \cite{genevey2008}, and the VDM reconstructions by \cite{genevey2008} and \cite{korte2005}). 
Here, we consider this set of reconstructions to estimate the uncertainty of the TSI reconstruction caused by uncertainties in the geomagnetic field.
}
\begin{table}
\caption{Paleogeomagnetic Models }           
\label{table3}      
\centering                                      
\begin{tabular}{l r r c}          
\hline\hline
Model & Reference & Period Covered & Concept	\\
\hline
KN08 	& (1)  &  9495 BC - 1705 AD &  VADM\\
GN08-8k & (2)  & 5995 BC - 1705 AD & VADM \\
GN08 & (2) & 995 BC - 1705 AD & VDM \\
KC05	         & (3)    & 4995 BC  - 1705 AD & VDM \\	
\hline
\end{tabular}
\tablebib{
(1)  \citet{knudsen2008};  (2)  \cite{genevey2008}; (3) \cite{korte2005}
}
\end{table}

\subsubsection{Reconstructions for the last 3,000 years}

\begin{figure*}
\centering
\includegraphics[width=13cm,clip=true, viewport=0.5cm 2cm 20cm 27cm]{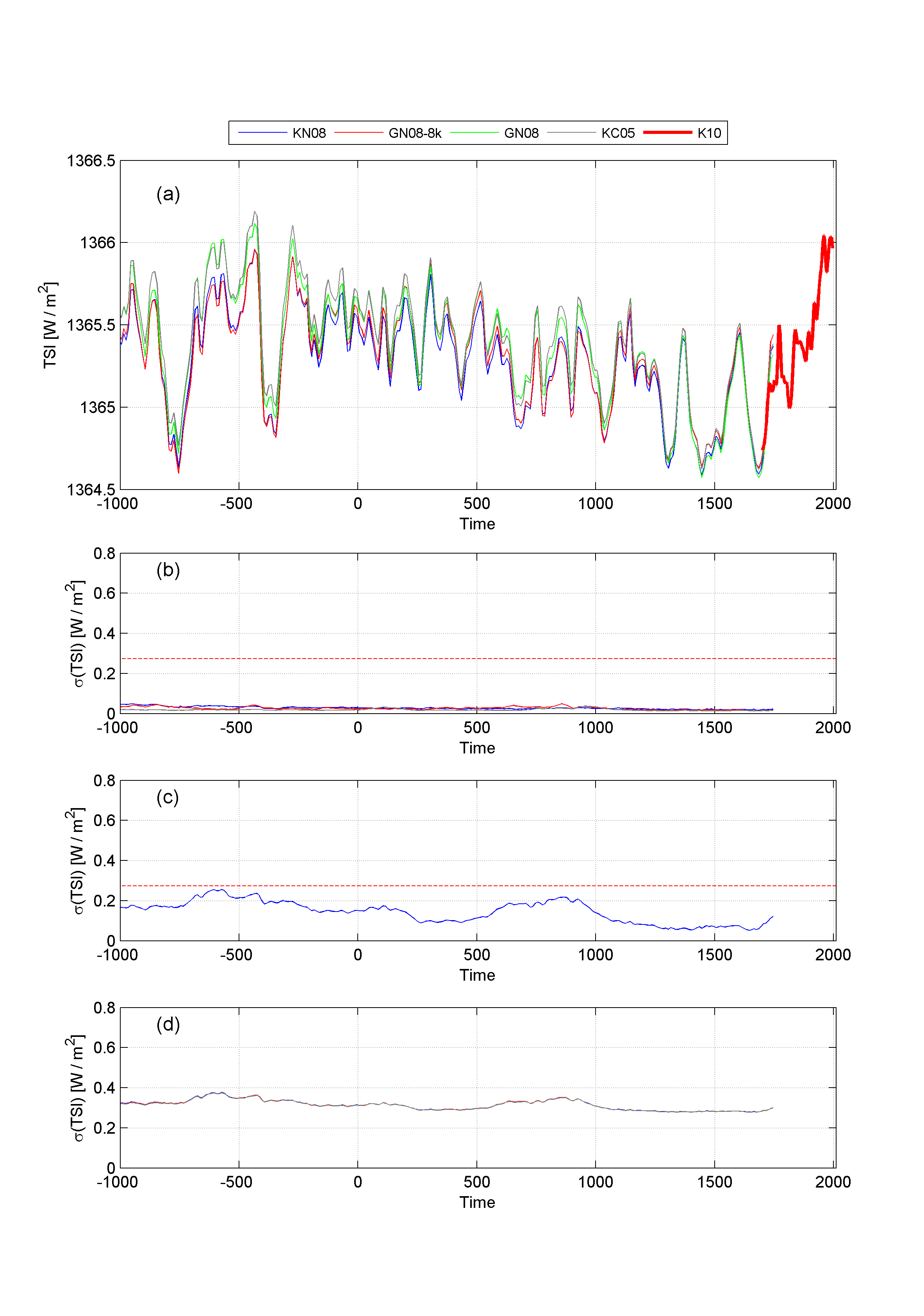}
\caption{(a) Reconstructions of the total solar irradiance for the last 3,000 years. All reconstructions are based on the INTCAL04 $\Delta$ \element[ ][14]{C} data set, but employing different reconstructions of the geomagnetic dipole momentum. The VADM reconstructions by \cite{knudsen2008} (KN08/blue line) and \cite{genevey2008} (GN08-8k/thin red line) are presented. In addition, the VDM reconstructions by \cite{genevey2008}(GN08/green line) and \cite{korte2005} (KC05/gray line) are also plotted. For reference, 10-year running means of the SATIRE-T reconstruction are displayed from 1700 to the present (K10/thick red line). (b) Estimate of the error due to the uncertainties of the open flux reconstructions. The red dotted line presents the uncertainty of the TSI computed using SATIRE-T. (c) Difference between the maximum and minimum values of the TSI computed employing different paleomagnetic models. (d) Estimated total uncertainties of the reconstructed TSI employing SATIRE-M model.}
\label{FigTsiLong_3k}
\end{figure*}

In Figure \ref{FigTsiLong_3k}, we present the reconstructions of the TSI for the last 3,000 years, which is the period
 over which all reconstructions of the geomagnetic field overlap. 
Panel (a) presents the reconstructions of the TSI. 
The reconstructions based on the VADM models by \cite{knudsen2008} and \cite{genevey2008} are presented as 
 blue and thin red lines, respectively. 
The reconstructions based on the VDM models by \cite{korte2005} and \cite{genevey2008} are displayed as the gray and cyan lines, respectively. 
This color scheme applies for panels (a), (b) and (d). For reference, the SATIRE-T reconstruction is presented from 1700 to present (thick red line). 
The different VADM models give results very similar to each other over the full plotted period of time, whereas the VDM models give
 somewhat higher irradiances, in particular the VDM reconstruction based on \cite{korte2005}, {because of their systematically lower
paleomagnetic dipole moment (see Fig. \ref{FigMagLong}, Sect.~\ref{Sec:geom} and Appendix \ref{Appen:stats})}.

Uncertainties in the reconstructions are presented in Fig. \ref{FigTsiLong_3k}b-d. 
Here, we take into account three major sources of uncertainties to the TSI computed from 10-year averaged values of open flux. 
The first one is the uncertainty in the TSI computed employing the SATIRE-T model ($\sigma_{TSI-SATIRE}$), which is represented
 by the red dotted lines in panels (b) and (c) and was discussed in Sect. 3.1. 
The second source of uncertainty lies in the computation of the open magnetic flux from cosmogenic isotopes ($\sigma_{\phi_{open}}$),
{which propagates into the TSI employing the procedure described in Sect. 2}. 
Panel (b) {presents} the estimate of the uncertainty of the $j$ data point due to the open flux
 ($\sigma_{TSI-\phi_{open}}^2 = (a_{TSI}\sigma_{\phi_{open}})^2 + (b_{TSI}\sigma_{\phi_{open}})^2$). 
The third source of uncertainty is the discrepancy between the reconstructions based on conceptually different
 paleomagnetic models ($\sigma_{TSI-GM}$). 
Panel (c) shows the difference between the maximum and minimum values of the $j^{th}$ TSI value computed employing
 different paleo-geomagnetic models ($\sigma_{TSI-GM}$). 
Panel (d) presents the total uncertainty that is given by the sum of the evaluated uncertainties
 ($\sigma_{total}^2 = \sigma_{TSI-SATIRE}^2 + \sigma_{TSI-\phi_{open}}^2 + \sigma_{TSI-GM}^2$). 
For this period, the major sources of uncertainty are related to the irradiance model itself and to the conceptually
 different paleomagnetic models. 
The uncertainty introduced by the open flux is much lower. 
Moreover, the overall uncertainty during this period does not exceed 0.4 W/m$^2$. 
{Note that we have followed common practice and have assumed that there were no significant changes in ocean mixing during the Holocene
 \citep{stuiver1991,usoskinkromer2005}.}

A summary of the statistical properties of the TSI reconstructions for the last 3000 years is plotted in Fig.~\ref{figure_tsi_long_histogram} of the appendix. From this figure and statistical tests we found that the distributions do not correspond to normal distributions. Additionally, reconstructions from the same group (VADM or VDM) have the same median values.  However, the reconstructions belonging to different groups disagree from each other (see Appendix \ref{Appen:stats}).

In the model, we assume that the network filling factor ($\alpha_n$) is proportional to the network magnetic flux ($\phi_{open} + \phi_{eph}$). The proportionality coefficient ($1/\phi_{sat,n}$) relates the physical properties of the network emission to the magnetic flux. However, it depends on the spatial resolution of the observations employed to compute it as well as on the estimate of the contribution of the bright components to the TSI. The SATIRE-T model suggests that the increase in the network emission since the Maunder Minimum is approximately 0.7 W/m$^2$. This value is close to the increase of 0.9 W/m$^2$ estimated by \cite{frohlich2009} and employed by \cite{steinhilber2009}. On the other hand, the reconstruction of the bright component of the TSI by \cite{wenzler2006} from magnetograms and continuum disk images indicates a contribution of approximately 0.2 W/m$^2$ during the minima between cycles 21-22 and 22-23. This comparison stresses the need for higher resolution and more sensitive observations of the solar surface magnetic field, such as now being provided by the observations by the Helioseismic and Magnetic Imager (HMI) instrument on board of the Solar Dynamics Observatory (SDO). Such data are expected to improve the estimate of the contribution of the bright components of the solar atmosphere to the TSI.

\subsubsection{Reconstructions for the Holocene}

\begin{figure*}
\centering
\includegraphics[width=13cm,clip=true, viewport=0.5cm 2cm 20cm 27cm]{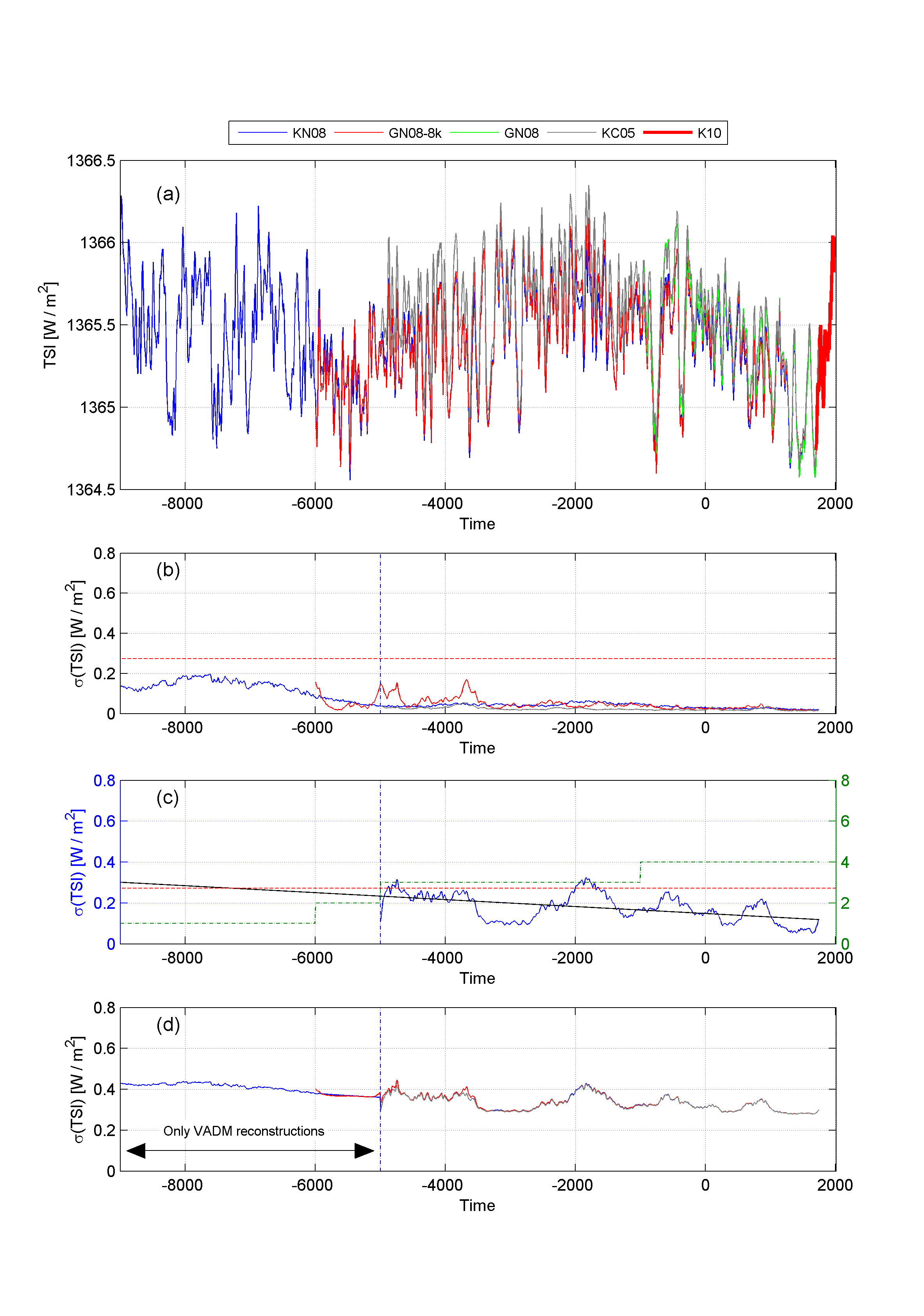}
\caption{Reconstructions of the total solar irradiance for the Holocene. The figure has the same structure as Fig. \ref{FigTsiLong_3k}, but for the period 9495 BC to the present. See the caption of Fig. \ref{FigTsiLong_3k} for more information. While the green line in panel (c) indicates the number of reconstructions employed to compute the uncertainty of the TSI due to the uncertainty of the evolution of the geomagnetic dipole moment, the black line represents the long-term trend of the uncertainty of the evolution of the geomagnetic dipole moment ($y = -1.7 \cdot 10^-5*t + 0.15$ [W/m$^2$]).  The period for which only VADM reconstructions are available is indicated in panel (d).}
\label{FigTsiLong}
\end{figure*}

Reconstructions of TSI for the whole Holocene are presented in Figure \ref{FigTsiLong}. The structure and color scheme are the same of the Figure \ref{FigTsiLong_3k}. The first 4,500 years are covered by the reconstructions based on the VADM values published by \cite{knudsen2008} and partially by the one based on the VADM presented by \cite{genevey2008}. VADM and VDM reconstructions are jointly available from approximately 5000 BC to 1700 AD, which is the period covered by the \cite{korte2005} reconstruction.

We evaluate the uncertainties for these reconstructions in the same way as we described in the previous section, by considering three major sources. The uncertainties of each reconstruction due to the uncertainties of the computation of the open flux are shown in panel (b). It is noticeable that the uncertainties are higher in the early part of the time series. {The} uncertainty estimated for the KN08 reconstruction decreases after approximately 6000 BC to the level of the other reconstructions.  In order to take into account the uncertainty related to the conceptually different paleomagnetic reconstructions, we consider separately the periods that VADM and VDM reconstructions are and are not jointly available. For the period that they are jointly available, i.e. after 5000 BC, we follow the method described in the previous section. For the period prior to jointly reconstructions, we linearly extrapolate the trends bound between 5000 BC and 1700 AD. The results are presented in panel (c). The dashed green line in this panel displays the number of reconstructions employed to compute the maximum and minimum values (scale on the right hand side). While the uncertainty due to the paleomagnetic models is on average at the same level as the uncertainty of the SATIRE-T model from 5000 BC to approximately 1000 BC, during the last 3000 years the uncertainty is lower. The total uncertainty for each reconstruction, which was computed as presented in Sect. 3.2.1, is shown in panel (d). The overall uncertainty is about 0.4 W/m$^2$.

As can be gathered from Fig. \ref{figure_tsi_tot_histogram} the TSI distributions on this longer time scale are significant closer to Gaussians. The difference to the more skewed distributions found for the last 3000 years has to do with the irregular distributions of grand minima. 

\begin{figure*}
\centering
\includegraphics[width=13cm,clip=true, viewport=0.5cm 2cm 20cm 27cm]{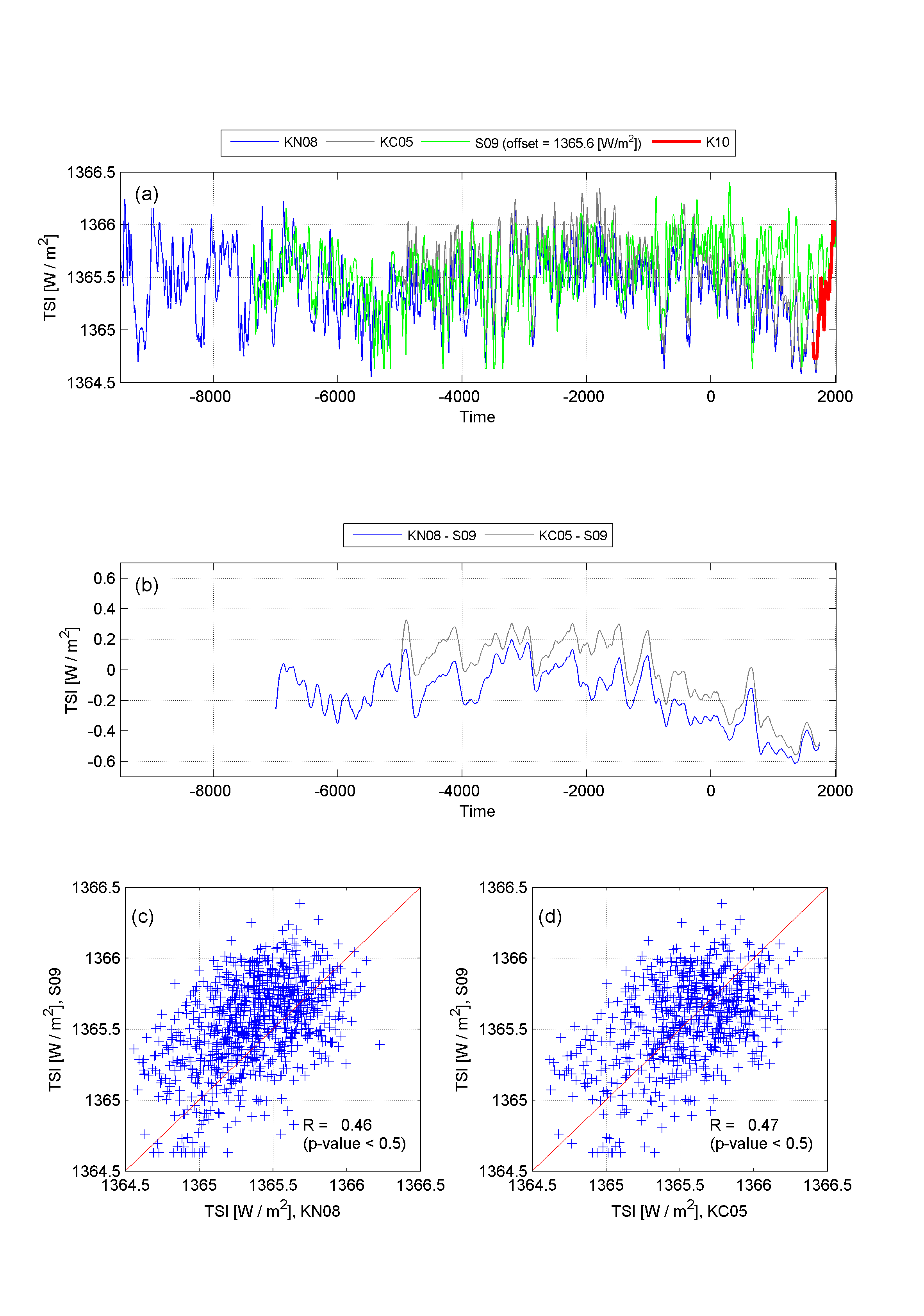}
\caption{(a) Comparison between the TSI reconstruction by \cite{steinhilber2009} (S09; VADM; green line) and two TSI reconstructions obtained in this paper based on VADM and VDM paleomagnetic reconstructions by \cite{knudsen2008} (KN08; VADM; blue line) and \cite{korte2005} (KC05; VDM; gray line), respectively. For reference, 10-year running means of the SATIRE-T reconstruction are displayed from aprroximately 1640 to the present (K10, thick red line). (b)  Differences between 120-year running means of the models based on \element[ ][14]{C} and \element[ ][10]{Be} records. (c) Scatter plot TSI-S09 versus TSI-KN08. (d) Scatter plot TSI-S09 versus TSI-KC05.}
\label{FigTsiLongS09}
\end{figure*}

\begin{figure*}
\centering
\includegraphics[width=13cm,clip=true, viewport=0.5cm 15cm 20cm 27cm]{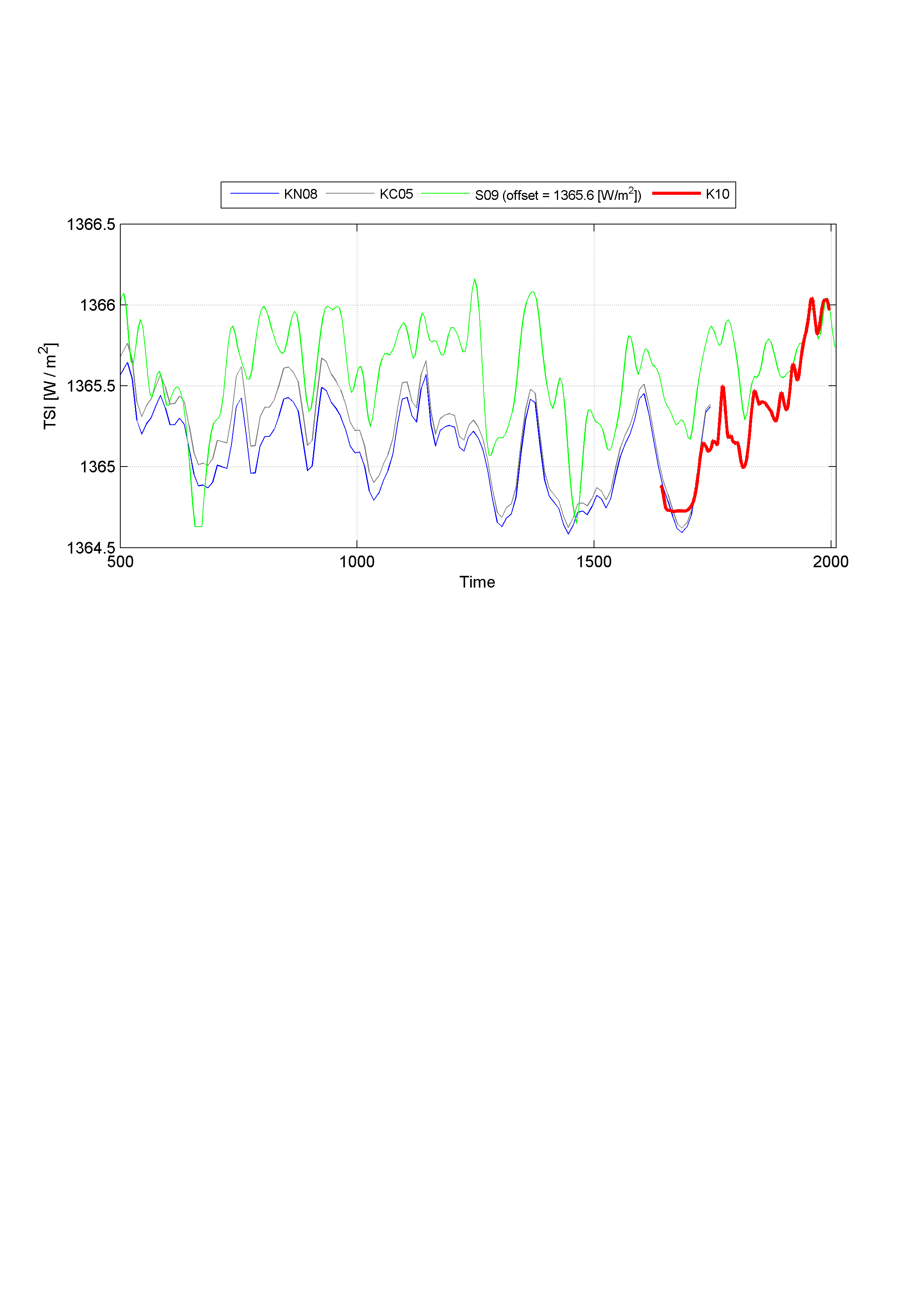}
\caption{Comparision between the TSI reconstructions based on \element[ ][14]{C} and \element[ ][10]{Be} from 500 AD to the present. Except for the considered time interval the structure of the figure is identical to panel (a) of Fig. \ref{FigTsiLongS09}.}
\label{FigTsiLongS09_zoom}
\end{figure*}

{\subsubsection{Comparison between TSI reconstructions based on \element[ ][14]{C} and \element[ ][10]{Be} records.}
\label{Sec:compBe}

In this section, we compare the TSI reconstruction based on \element[ ][10]{Be} records presented by \cite{steinhilber2009} and two TSI reconstructions based on \element[ ][14]{C} computed in this paper. Figure \ref{FigTsiLongS09}a shows the TSI reconstructions by \cite{steinhilber2009} (S09; VADM; green line) and the reconstructions based on the \cite{knudsen2008} (KN08; VADM; blue line) and \cite{korte2005} (KC05; VDM; gray line) paleomagnetic models since approximately 9500 BC. For reference, 10-year running means of the SATIRE-T reconstruction are displayed from approximately 1640 AD to the present (K10, thick red line). The S09 reconstruction, which was obtained from 40-year running means of \element[ ][10]{Be}, was resampled to decadal values employing a linear interplolation. An offset of 1365.6 W/m$^2$ was introduced to match the present value of the TSI (PMOD composite). 

{The TSI reconstructions based on different approaches are in the same variability range, while producing some discrepancies at mid to long-term time scales.} Although the values of the three time series lie in the same interval and display similar trends prior to 1000 BC, the reconstructions based on \element[ ][14]{C} and \element[ ][10]{Be} diverge during the last 3000 years. Figure \ref{FigTsiLongS09}b offers a better visualization of the structural difference between the time series. In this figure, the differences between 120-year running means of the models based on \element[ ][14]{C} and \element[ ][10]{Be} records are presented. For the period analyzed, the standard deviation of the differences KN08-S09 and KC05-S09 are approximately 0.32 and 0.34 W/m$^2$, respectively. {Note that the differences between \element[ ][14]{C} and \element[ ][10]{Be} reconstructions are larger than the differences due to the uncertainty of the geomagnetic field.}  Panels (c) and (d) of Fig. \ref{FigTsiLongS09} show the scatter plots of the S09 reconstruction versus the KN08 and KC05 reconstructions, respectively. The correlation coefficients are indicated in the panels. As a consequence of the structural differences of the time series, the correlation coefficients between the reconstructions based on \element[ ][14]{C} and \element[ ][10]{Be} records are relatively low. 

During the last 3000 years the S09 reconstruction suggests a higher level of TSI compared to the KN08 and KC09 reconstructions. For a better visualization, we present in Fig. \ref{FigTsiLongS09_zoom} the evolution of the time series from 500 AD to the present. Near the end of the Maunder Minimum, the difference is approximately 0.5 W/m$^2$ if we require the TSI reconstruction by \cite{steinhilber2009} to match the present value of the TSI. Since the two reconstructions are based on different approaches and cosmogenic isotopes, it is difficult to identify the major source of the discrepancy. Different long-term trends in the \element[ ][14]{C} and \element[ ][10]{Be} data certainly play an important role \citep{vonmoos2006, usoskin2009}, but methodological differences may also contribute. Note that the TSI reconstructed by S09 also shows a much flatter trend than that obtained by \cite{krivova2010a} from telescopically measured group sunspot number (K10). }

\subsubsection{Discussion of examples of reconstructed irradiance trends}

\begin{figure*}
\centering
\includegraphics[width=13cm,clip=true, viewport=0.5cm 15cm 20cm 27cm]{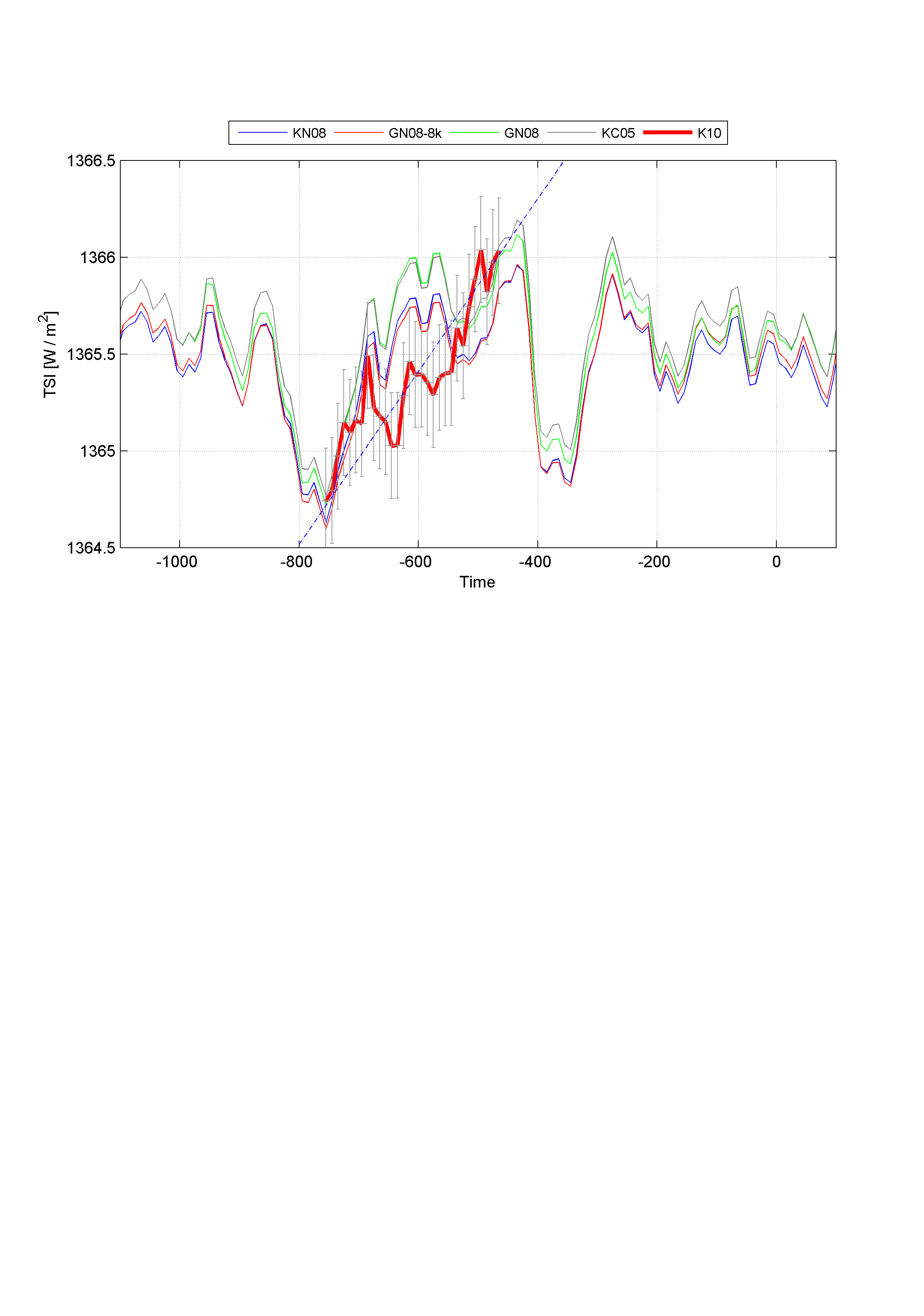}
\caption{Comparison between the evolutions of the TSI after the grand minimum observed in approximately 765 BC and after the Maunder Minimum.  The figure follows the format of Fig.  \ref{FigTsiLong_3k}a. The 10-year averaged SATIRE-T reconstruction (thick red line) is shifted in time by 2465 years in order to match the minimum value observed in 765 BC. The dashed blue line connects the beginning and the end of the SATIRE-T reconstruction and emphasizes the roughly linear rise in irradiance sind the end of the Maunder Minimum.}
\label{FigTsiLong_3k_zoom}
\end{figure*}

The reconstructions indicate that approximately 2450 years ago the TSI was at the same level as is observed at the present time. Figure \ref{FigTsiLong_3k_zoom} shows the TSI reconstructions from 1100 BC to 100 AD. According to the VDM and VADM reconstructions, the peak value ranges from approximately 1366.0 to 1366.2 W/m$^2$, respectively. This peak was reached about 330 years after the {\bf preceding} grand minimum, which was observed in 765 BC. The TSI values during this grand minima computed employing different paleomagnetic reconstructions ranges from 1364.6 to 1364.8 W/m$^2$. This grand minimum was previously discussed by \cite{eddy1977a, eddy1977b}, \cite{stuiver1980}, \cite{stuiver1989}, \cite{goslar2003} and \cite{usoskin2007}. For reference, the decadally averaged  SATIRE-T reconstruction (thick red line) is shifted 2455 years to match the minimum value observed in 765 BC. The dashed blue line connects the beginning and the end of the SATIRE-T reconstruction. The amplitude and time scale of the TSI evolution from the previous grand minimum, which occurred at approximately 765 BC, to the grand maximum is similar to the amplitude and time scale reconstructed from the end of Maunder Minimum to the present. However, there are also noticeable differences of the evolution of the two periods. For roughly 200 years after the 765 BC grand minimum the TSI is higher than for the same interval after the Maunder Minimum. This maximum around 430 BC was followed by the next grand minimum, which is seen as a large and very rapid decrease of the TSI. This extended minimum lasted for approximately 150 years. The minimum value of the TSI, which ranges from 1364.8 to 1365.0 W/m$^2$, is reached in 345 BC. Note that all 3 plotted grand minima have the same TSI level within the uncertainty limits. The similarities between the evolution of the TSI for the ancient and the present periods cannot be used to predict the occurrence of a grand minimum in the near future, but shows that very rapid and deep drops in the irradiance can occur starting from a high level of TSI, such as the present.

\begin{figure*}
\centering
\includegraphics[width=13cm,clip=true, viewport=0.5cm 15cm 20cm 27cm]{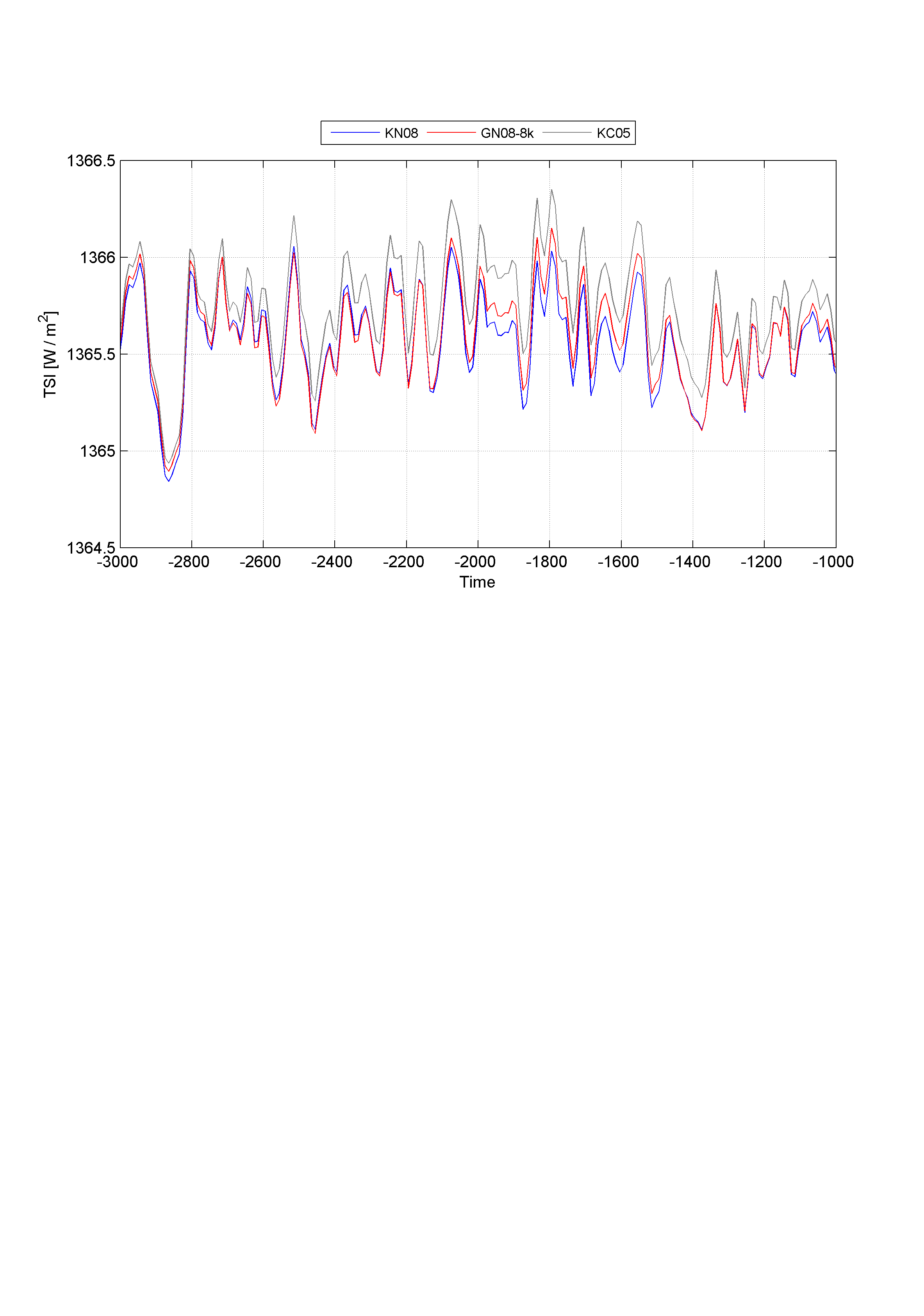}
\caption{Evolution of the TSI during the period of high solar activity from 3000 BC to 1000 BC. The figure follows the format of Fig.  \ref{FigTsiLong_3k}a. }
\label{figure_tsi_long_2000BC}
\end{figure*}

The reconstructions also suggest that from about 3000 BC to 2000 BC the peaks in solar irradiance exhibited values at a similar level as observed at the present time. Figure \ref{figure_tsi_long_2000BC} shows the reconstructions and uncertainties from 3000 BC to 1000 BC. As discussed in the previous subsection, the TSI reconstruction based on the VDM reconstruction \citep{korte2005} displays higher values, which are for some periods slightly higher than the averaged value observed during the last three solar cycles. The structure of the signal during this period is clearly different from the episode described in the previous paragraph. With the exception of a period close to 2860 BC, pronounced grand minima are not observed, although large amplitude oscillations are present. Obviously, the TSI exhibited a rather different behavior during that period of time than in the last 400 years.

\begin{figure*}
\centering
\includegraphics[width=13cm,clip=true, viewport=0.8cm 2cm 20cm 27cm]{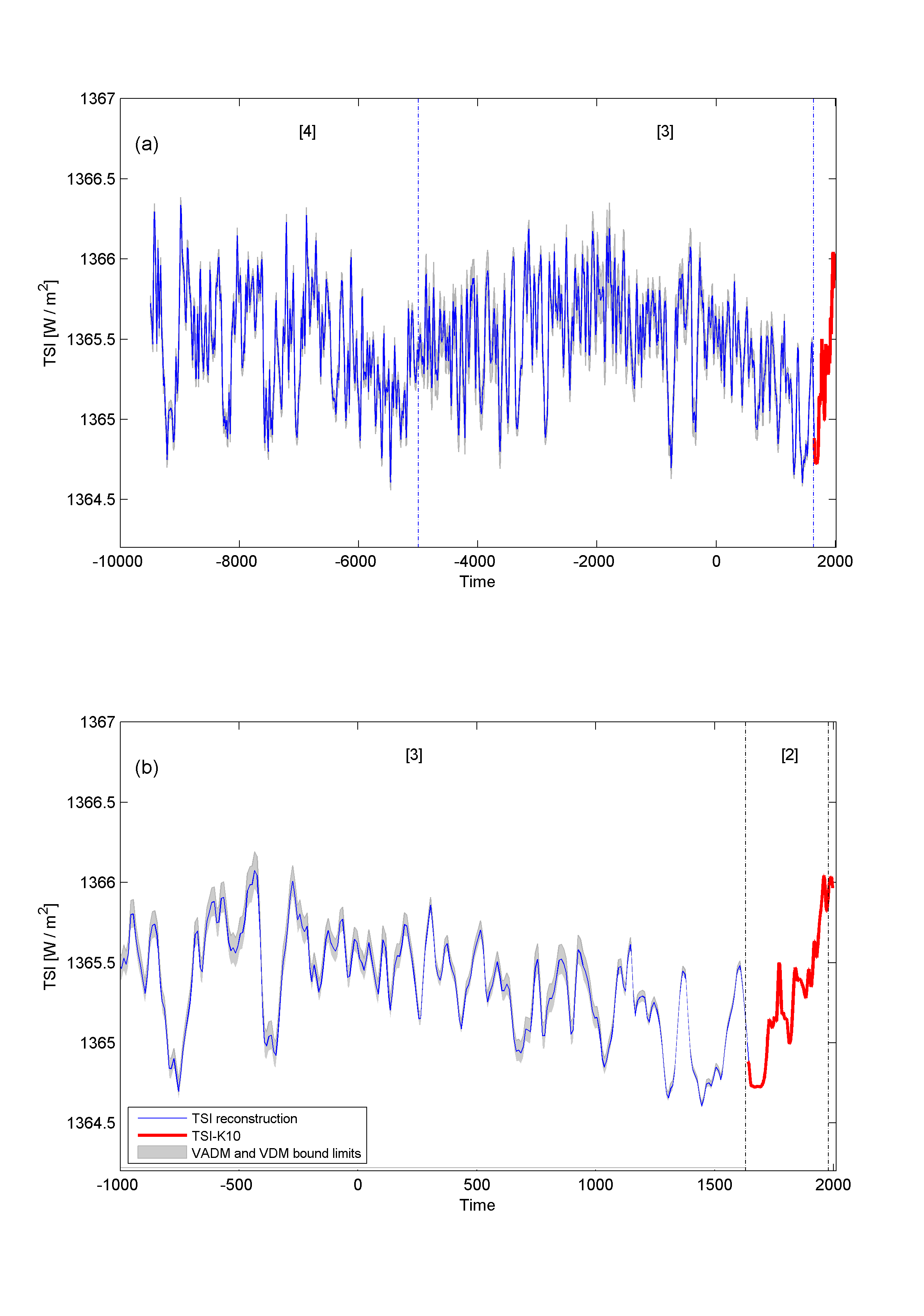}
\caption{TSI weighted reconstruction since approximately 9500 BC. In order to provide a better visualization, the evolution since 1000 BC is displayed in panel (b). The filled gray band represents region limited by the KN08-VADM and KC05-VDM reconstructions. 
For reference, the red lines represent the 10-year averaged reconstruction by \cite{krivova2010a}.}
\label{figure_tsi_ensamble}
\end{figure*}

\subsection{Construction of a TSI time series for investigations of the climate evolution during the Holocene}

One of the main motivations for reconstructing the TSI over the Holocene is to produce a consistent record of the external heat input into the {coupled} atmosphere/ocean/land system that is needed to investigate the evolution of the Earth's climate employing general circulation models \citep{schmidt2010,schmidt2010a}. However, in the previous sections we have produced multiple TSI time series whose validity and length depends on that of the particular geomagnetic field {reconstruction} underlying it. For use with climate models, however, it is desirable to construct an optimal time series extending through the {whole} Holocene. In this subsection, we describe a procedure {for combining} the SATIRE reconstructions during different periods of time. This is not completely trivial due to the different lengths of the time series we are averaging and the fact that they partly have different median values.

{In order to produce a representative time series, we divide the Holocene into four periods according to the available data for each period. Table \ref{table5} summarizes the periods considered and the data available for each period.  After the end of the Maunder Minimum we compute daily values, while prior to the end of the Maunder Minimum we compute 10-year averages. For the period for which both solar disk magnetograms and continuum images are available (period 1) we employ the SATIRE-S reconstruction. SATIRE-T reconstruction is used from the beginning of the Maunder Minimum ($\sim$1640 AD) to 1978 AD. Prior to 1640 AD reconstructions based on cosmogenic isotopes are used. As VDM paleomagnetic reconstructions are available since approximately 5000 BC, we have to treat separately periods 3 and 4. We emphasize that the reconstructions based on different proxies have different time resolutions.

For period 3, VADM and VDM reconstructions are available. The TSI reconstructions based on conceptually different paleomagnetic models differ significantly from each other. As discussed in Sect 3.2, the paleomagnetic reconstructions based on the VADM and VDM models provide the upper and lower limits of the evolution of the geomagnetic field, respectively. Consequently, we need to combine them to produce a representative reconstruction for this period. As we have discussed previously, the median values of the reconstructions for the last 3000 years that belong to a given group are not significantly different. In this way, assuming that this result holds for the whole period starting in 5000 BC, we can compute the final TSI at time step $j$ as an average
of the two methods VADM and VDM taking into account that they represent lower (VDM) and upper (VADM) bounds to the true dipole moment. We do not claim that the average of the two groups is the true value, but since the true value lies between these bounds, if (the true value) is more likely to lie close to the average value than to either of the bounds. Here we employ the reconstructions based on the KC05 and KN08 dipole moments that are representative of the VDM and VADM groups, respectively.

For period 4, only VADM reconstructions are available. Here we employ the KN08 reconstruction as the reference. To be consistent with other periods, it is necessary to introduce an offset that can be estimated from the difference of the mean values of the KN08-VADM and KC05-VDM reconstructions for the period 3. {In this way, we implicitly assume that the VDM paleomagnetic models would have lower values than the VADM paleomagnetic models in period 4 as they have had in more recent times as well.}
The numerical value of the offset is approximately 0.05 W/m$^2$. In this way, discontinuities are not introduced in the interface between periods 3 and 4. {Note that the upper bound limit in Fig. \ref{figure_tsi_ensamble} is obtained by adding $2 \cdot offset$ to the VADM reconstruction}. For more recent times the SATIRE-T and SATIRE-S based reconstructions are appended to the \element[ ][14]{C} based TSI reconstructions (see Table \ref{table5}). The final TSI reconstruction is presented in Figure \ref{figure_tsi_ensamble}. The periods considered and listed in Table \ref{table5} are indicated in the figure. To avoid overloading the figure, 10-year averages of the SATIRE-T model are plotted.}

\begin{table*}
\caption{ SATIRE Models employed to construct a representative TSI time series for the Holocene.}
\label{table5}
\centering
\begin{tabular}{clllll}
\hline
Index & Period &  Cadence & Model & Input Data\\
\hline\hline
1 & 1974 AD - 2010 & Daily & SATIRE-S & Disk magnetograms and continuum images &  \\
2 & 1640 AD - 1974 AD & Daily  & SATIRE-T & Sunspot observations  &  \\
3 & 5000 BC - 1640 AD  & Decadal  & SATIRE-M & Cosmogenic isotopes ( \element[ ][14]{C};  VADM and VDM geomagnetic field) \\
4 & 9495 BC - 5000 BC   &  Decadal & SATIRE-M & Cosmogenic isotopes ( \element[ ][14]{C}; VADM geomagnetic field) \\
\hline
\hline
\end{tabular}
\end{table*}


\section{Concluding remarks}

In order to understand the evolution of the Earth's climate during the Holocene the variability of the internal and external drivers needs to be quantified. Among the external drivers, the solar irradiance plays a prominent role. Here, we have reconstructed the TSI over the Holocene. This is an extension of the SATIRE family of models to a longer time scale. It represents the first reconstruction of TSI during pre-telescopic times {that is based on a physically consistent treatment, which builds on} the assumption that the magnetic field is responsible for all TSI changes. This assumption implies that our model cannot handle changes in TSI that do not originate in surface magnetism (e.g. convective efficiency changes unrelated to surface magnetism or effects associated with {\it r} modes, i.e. oscillatory modes related with rotation, or changes associated with magnetic fields buried {deep} in the convection zone). {On the other hand, it allows us to rigorously derive the proper relationship between irradiance and magnetic flux.} Earlier reconstructions {of TSI based on multi-decadal averages of cosmogenic data} \citep[e.g.][]{steinhilber2009} are based on linear extrapolations of an empirical relationship between open flux and TSI. However, our computations do not fully support a simple one-to-one relationship between TSI and solar open magnetic flux {on decadal time-scale}, since {we find that} magnetic flux from two cycles contributes to the TSI from one cycle. {The reconstructions carried out here are based on this consistently derived relationship. Another major difference between the present and previous reconstructions of TSI over the Holocene lies in our careful assessment of the uncertainties in reconstructions of the geomagnetic field by considering the available geomagnetic field reconstructions that reaches back at least 3000 years. The various reconstructions of the geomagnetic field can lead to significant differences in TSI at earlier times. }

The measured solar irradiance can be reproduced on the {basis of the} assumption that its variability is related to the evolution of the magnetic features lying on the solar surface \citep{krivova2003, wenzler2006}. However, the spatial and temporal resolution of records of magnetic features on the solar disk degrades progressively as we go back in time. For example, the solar irradiance can be reproduced with high accuracy from magnetograms and continuum images of the solar disk employing a physics-based model \citep[SATIRE-S][]{krivova2003}. However, these records are jointly available only for the last three solar cycles. To deal with this limitation, models of the evolution of the solar magnetic field can be employed to replace the direct observations of the evolution of the magnetic features on the solar disk. Most of these models are based on the records of sunspot number, which are available since approximately 1610 AD (e.g., SATIRE-T \citep{balmaceda2007, krivova2007, krivova2010a} and other models \citep{lean1995, solanki1999, wang2005, tapping2007, crouch2008}), but contain no information on the spatial distribution of the magnetic features on the solar surface and suffer from loss of temporal resolution at earlier times. On longer time-scales, the evolution of the solar magnetic flux can be evaluated based on records of cosmogenic isotopes stored in natural archives, although generally at the cost of a further loss of temporal resolution. Here, we have shown that 10-year averages of the total and spectral irradiance can be obtained by employing the magnetic flux computed from cosmogenic isotopes as the input of the SATIRE model (SATIRE-M, where 'M' stands for millennial time scales). Note that the core of the irradiance reconstruction method is the same for all variations of the SATIRE model. The solar atmosphere is divided into components: sunspots, umbrae and penumbrae, faculae, network and the quiet Sun. The evolution of the spectrum is described by the filling factors on different parts of the solar surface of each component and its time independent brightness. Only the evolution of the filling factors of the surface components is computed according to the available data.


The reconstructed 10-year averaged TSI displays variations over a range of approximately 1.5 W/m$^2$. This is slightly larger than the 1.3 W/m$^2$ between the Maunder Minimum and the current grand maximum found by \cite{krivova2010a}, mainly because an earlier Grand minimum (around 7500 BC) displays a lower TSI than the Maunder Minimum and the Grand Maximum at the beginning of the time series displays a larger TSI than the just ending Grand Maximum. We point out that this range is uncertain as there are significant uncertainties in the computation of the open flux at the earlier times. The major sources of uncertainty of the solar irradiance reconstructions during the Holocene come from the simplifications employed to derive the evolution of the magnetic flux emerging in AR and ER based on decadal values of the open flux and from the evolution of the Earth's magnetic field. In order to evaluate the uncertainties due to the evolution of the Earth's magnetic field, we compared an ensemble of reconstructions that are based on two conceptually different approaches to computing the evolution of the Earth's magnetic dipole. For the period that all reconstructions are jointly available (i.e. during the last 3,000 years) the two groups show significantly different median values, reflecting the fact that reconstructions of the virtual dipole moment may underestimate the true dipole and, consequently, overestimate the solar parameters. The opposite is expected for the virtual axial dipole moment reconstructions. In addition to these uncertainties, we recall that the reconstructions are based on the assumption that carbon cycle parameters have not changed significantly throughout the Holocene, which may not be exactly valid, particularly in the early Holocene, as discussed previously by \cite{solanki2004}, \cite{ellison2006}, \cite{usoskin2007,usoskin2009}, and others. {We stress that the TSI reconstructions presented in this paper are based on  one individual record of \element[ ][14]{C}, and, as shown in Sect. \ref{Sec:compBe}, using \element[ ][10]{Be} records would probably lead to another long-term behaviour.}

We have also produced a reconstruction that is a careful combination of the various TSI reconstructions we have carried out for the period prior to the Maunder Minimum, i.e. before 1640 AD. Between 1640 AD and 1974 AD it corresponds to the SATIRE-T reconstruction of \cite{krivova2010a}, while after that we employ a SATIRE-S reconstruction based on Kitt Peak \citep{wenzler2006} and MDI \citep[following][]{krivova2003}  magnetograms. This combined reconstruction, which is our best estimate of the TSI during the Holocene,  is available as electronic material to this paper and can also be downloaded from the website: http://www.mps.mpg.de/projects/sun-climate/data.html.

The present analysis is in agreement with earlier reconstructions of the solar activity during the Holocene regarding the occurrence of extended periods of low and high solar activity \cite[e.g.][]{usoskin2007}. These coincide with periods of low and high TSI. We noted some similarities between the evolution of the solar activity from the Maunder Minimum to the present and an episode observed approximately 2800 years ago. Both episodes displayed changes of the solar irradiance of about 1.3 W/m$^2$ in an interval of approximately 300 years. The ancient episode ended rather abruptly in a grand minimum, which emphasizes that a very rapid drop in the irradiance can occur starting from a high level of TSI, such as the present one. However, we stress that we cannot employ these similarities to predict the occurrence of a grand minimum in the near future.

{\begin{appendix}
\section{Statistical properties of the TSI reconstructions}
\label{Appen:stats}

Panels (a) to (d) of Fig.  \ref{figure_tsi_long_histogram} display the probability distributions of the 10-year averaged TSI values. 
Each panel shows the probability distribution functions (PDF) for the normal distribution computed employing the respective mean
 and standard deviation values that are also given in each panel. 
In order to verify if the reconstructed TSI values are normally distributed, we performed a test of the default null hypothesis
 that the values come from a normal distribution, against the hypothesis that they do not come from this distribution. The test statistic is
\begin{equation}
T = max \left[ abs \left (ecdf(TSI) -ncdf(TSI) \right ) \right ]\\,
\end{equation}
where $ecdf(TSI)$ is the empirical cumulative distribution function (CDF) estimated from the TSI distribution 
 and $ncdf(TSI)$ is the normal CDF with mean and standard deviation equal to the mean and standard deviation of the TSI distribution. 
Recall that the value of the ecdf at a given value of the TSI is obtained from the PDF by integrating the latter from its lower
 end up to that particular TSI value. 
The $T$ values are indicated in the respective panels. 
The null hypothesis is rejected at a given significance level when $T$ is greater than a critical value. 
The test employed here uses a table of critical values computed using Monte-Carlo simulations, where the critical value for the significance level of 0.05 is 0.0547. 
We found that the null hypothesis of all reconstructions is rejected at this significance level, i.e., the distributions
 do not correspond to normal distributions. 
This is supported by a visual inspection which reveals the strong asymmetry in all the the empirical distributions. 
Besides the main peak located above 1365 W/m$^2$ there is also a lower (and statistically less significant) secondary
 peak below 1365 W/m$^2$, which corresponds to the relatively common grand minima in this period of time.

Figure \ref{figure_tsi_long_histogram}e compares the median values (red lines). 
Additionally, the upper and lower boundaries of the boxes display the respective quartile values. 
The dashed lines show the regions extending from the boundaries of each box to the most extreme values within
 1.5 times the interquartile range. 
As expected, the VDM reconstructions display higher median values than the VADM reconstructions. 
Furthermore, as also suggested by the empirical CDFs, the VDM distributions are shifted in relation to
 the VADM distributions, and also display a larger range of irradiance values.

We employ the Wilcoxon test to verify if two reconstructions come from identical distributions with
 equal medians, against the alternative that they do not have equal medians. 
Note that the distributions need not necessarily be normal. 
{Table \ref{table4} summarizes the results.} 
The reconstructions from the same group (VADM or VDM) have the same median values. 
However, the reconstructions belonging to different groups disagree from each other.

The statistical properties of the TSI reconstructions for the whole Holocene based on the paleomagnetic reconstructions by \cite{knudsen2008},  
 \cite{genevey2008}, and \cite{korte2005} are presented in panels (a)-(c) of Fig. \ref{figure_tsi_tot_histogram}, respectively. 
As in Fig. \ref{figure_tsi_long_histogram}, the red lines represent the probability distribution functions for the normal distribution. 
Note that each time series covers a different time span (see Table \ref{table3}). 
Tests of the null hypothesis that the values come from a normal distribution, as discussed in Sect. 3.2.1,
 indicate that it is not possible to reject this hypothesis at the significance level of 0.05 for the reconstructions based on 
 the \cite{genevey2008} paleomagnetic reconstruction. 
A visual inspection shows that the reconstruction based on the paleomagnetic model by \cite{korte2005}, that
 extends from 5000 BC to 1700, gives a slightly asymmetrical distribution. 
This behavior reflects the more frequent occurrence of grand minima in the later part of the time series, as can be gathered by comparing with the far more asymmetric distributions in Fig. \ref{figure_tsi_long_histogram}.

\begin{table*}
\caption{ Wilcoxon rank sum test of the distribution medians.}
\label{table4}
\centering
\begin{tabular}{l c c c c c c}
\hline
 & KN08 & GN08-8k& GN08& KC05\\
\hline\hline
KN08 & X &   & rej. & rej.\\
\hline
GN08-8k &  & X & rej. & rej. \\
\hline
GN08 &rej. & rej. & X & \\
\hline
KC05 & rej. & rej. & & X \\
\hline
\hline
\end{tabular}
\end{table*}

\begin{figure*}
\centering
\includegraphics[width=13cm,clip=true, viewport=0.5cm 8cm 19cm 27cm]{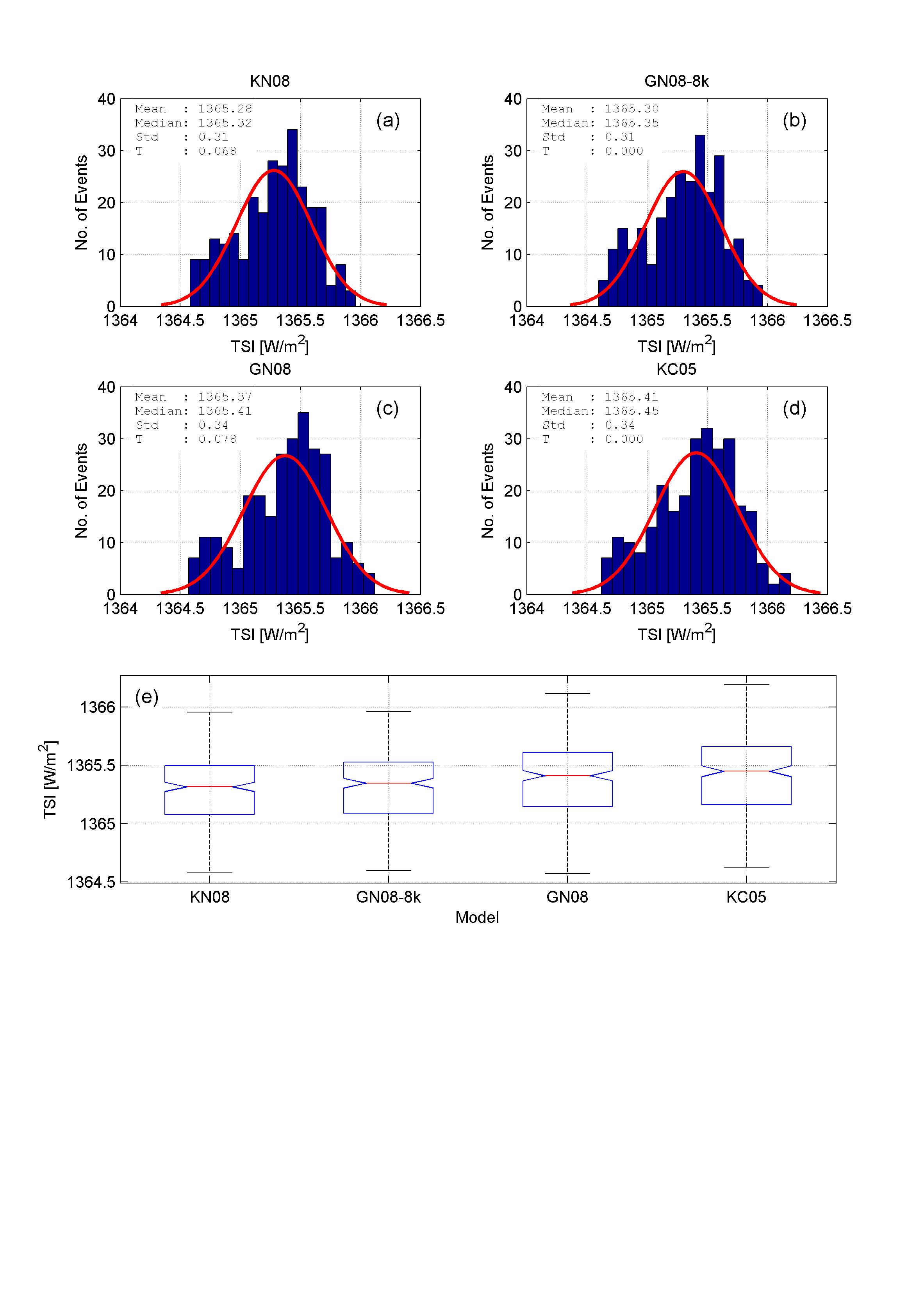}
\caption{Comparison between the statistical properties of the TSI reconstructions for the period 1000 BC to 1700 AD. Panels (a) to (d) present the distributions of the TSI reconstructions. The mean, median and standard deviation values are indicated in the panels. The red lines represent the probability distributions functions for the normal distribution. Panel (e) allows a visual comparison between the distributions. The median (red line), lower and upper quartile values of the distributions (lower and upper ends of the blue boxes) are plotted in each box. The black lines extend from the end of each box to the most extreme values within 1.5 times the interquartile range. Note that the interquartile range, which is also a measure of the statistical dispersion of the data,  is equal to the difference between the third and first quartiles.}
\label{figure_tsi_long_histogram}
\end{figure*}

\begin{figure*}
\centering
\includegraphics[width=13cm,clip=true, viewport=01.5cm 0.1cm 19cm 10cm]{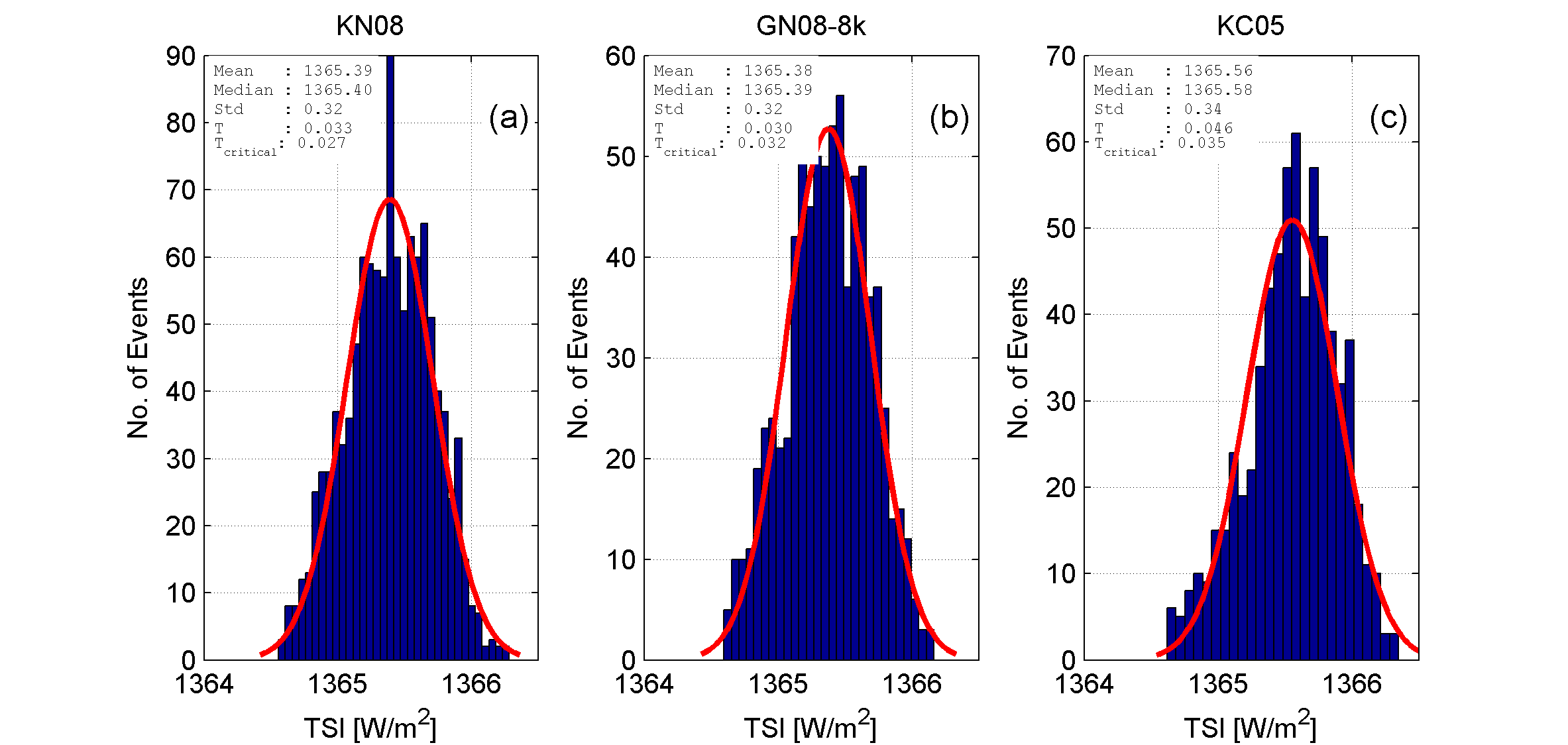}
\caption{Comparison between the statistical properties of the TSI reconstructions during the Holocene. The time coverage of each time series is presented in Table \ref{table3}. Panels (a) to (d) present the distributions of the TSI reconstructions. The red line represents the normal distribution computed employing the sample mean and standard deviation values of as the overplotted sample (the values are indicated in each panel). Additionally, the test statistic ($T$) and the critical ($T_{critical}$) values are listed.}
\label{figure_tsi_tot_histogram}
\end{figure*}

\end{appendix} 
}

\begin{acknowledgements}
We are grateful to C. Fr\"ohlich and F. Steinhilber for providing the PMOD composite TSI and the TSI reconstruction based on \element[ ][10]{Be} records, respectively. We are also thankful to M. Lockwood for providing the open flux reconstructions based on the aa-index. This work was supported by the Deutsche Forschungs-gemeinschaft, DFG project number SO 711/1-2, by the WCU grant No. R31-10016
funded by the Korean Ministry of Education, Science and Technology, and by the European Commission's Seventh Framework Programme (FP7/2007-2013) under
the grant agreement n° 218816 (SOTERIA project).
\end{acknowledgements}

\bibliography{TSIversion08}{}
\bibliographystyle{aa} 

\end{document}